\documentclass[prc, amsfonts, amssymb, amsmath, nofootinbib, showkeys, singlecolumn, superscriptaddress, notitlepage, preprintnumbers]{revtex4-2}
\usepackage{enumitem}
\usepackage{epsfig}
\usepackage{bm}
\usepackage{amssymb}
\usepackage{amsmath}
\usepackage{color}
\usepackage{dcolumn}
\usepackage{tensor}

\usepackage{svg} 

\usepackage[utf8]{inputenc}

\usepackage[normalem]{ulem}

\usepackage{amsmath}
\usepackage{amssymb}
\usepackage{amsthm} 
\usepackage{amscd, amsfonts, mathrsfs}
\usepackage{cases}
\usepackage{verbatim} 
\usepackage{epsf}
\usepackage{xcolor}
\usepackage[bookmarksnumbered,pdfpagelabels=true,plainpages=false,colorlinks=true,
   linkcolor=black,citecolor=black,urlcolor=black]{hyperref}
\usepackage{enumitem}

\usepackage{empheq}

\theoremstyle{plain}
\newtheorem{theorem}{Theorem}

\newtheorem{proposition}{Proposition}
\newtheorem{corollary}[theorem]{Corollary}

\theoremstyle{definition}

\allowdisplaybreaks

\newcommand{\Umb}{\textrm{\textbf{U}}}

\newcommand{\be}{\begin{equation}}
\newcommand{\ee}{\end{equation}}
\newcommand{\bea}{\begin{eqnarray}}
\newcommand{\eea}{\end{eqnarray}}

\newcommand{\bml}{\begin{subequations}}
\newcommand{\eml}{\end{subequations}}

\newcommand{\bbm}{\begin{bmatrix}}
\newcommand{\ebm}{\end{bmatrix}}
\newcommand{\bvm}{\begin{vmatrix}}
\newcommand{\evm}{\end{vmatrix}}

\colorlet{kigreen}{green!60!black}

\begin{document}
\title{Nonlinear causality of Israel-Stewart theory with diffusion}
\date{\today}

\author{Ian Cordeiro}
\email{itc2@illinois.edu}
\affiliation{Illinois Center for Advanced Studies of the Universe\\ Department of Physics, 
University of Illinois Urbana-Champaign, Urbana, Illinois 61801, USA}

\author{F\'abio S.\ Bemfica}
\email{fabio.bemfica@ufrn.br}
\affiliation{Department of Mathematics, Vanderbilt University, Nashville, Tennessee, 37211, USA}
\affiliation{Escola de Ci\^encias e Tecnologia, Universidade Federal do Rio Grande do Norte, Rio Grande do Norte, 59072-970, Natal, Brazil}

\author{Enrico Speranza}
\email{enrico.speranza@cern.ch}
\affiliation{Department of Physics and Astronomy, University of Florence, Via G. Sansone 1, 50019 Sesto Fiorentino, Italy}
\affiliation{Theoretical Physics Department, CERN, 1211 Geneva 23, Switzerland}

\author{Jorge Noronha}
\email{jn0508@illinois.edu}
\affiliation{Illinois Center for Advanced Studies of the Universe\\ Department of Physics, 
University of Illinois Urbana-Champaign, Urbana, IL 61801, USA}

\preprint{CERN-TH-2025-143}
\begin{abstract}
We present the first fully nonlinear causality constraints in $D = 3 + 1$ dimensions for Israel-Stewart theory in the presence of energy and number diffusion in the Eckart and Landau hydrodynamic frames, respectively. These constraints are algebraic inequalities that make no assumption on the underlying geometry of the spacetime or the equation of state. In order to highlight the distinct physical and structural behavior of the two hydrodynamic frames, we discuss the special ultrarelativistic ideal gas equation of state considered in earlier literature in $D = 1 + 1$ dimensions, and show that our general $D = 3 + 1$ constraints reduce to their results upon an appropriate choice of angles. For this equation of state in both $D = 1 + 1$ and $D = 3 + 1$ dimensions one can show that: (i) there exists a region allowed by nonlinear causality in which the baryon current transitions into a spacelike vector in the Landau frame, and (ii) an analogous argument shows that the solutions of the Eckart frame equations of motion never violate the dominant energy condition, assuming nonlinear causality holds. We then compare our results with those from linearized Israel-Stewart theory and show that the linear causality bounds fail to capture the new physical constraints on energy and number diffusion that are successfully obtained through our nonlinear causality approach.
\end{abstract}

\maketitle


\section{Introduction}

Relativistic viscous hydrodynamics \cite{Rocha:2023ilf} describes the dynamics of various physical systems, including the quark-gluon plasma  \cite{Heinz:2013th,Gale:2013da,Romatschke:2017ejr}, neutron star mergers \cite{Baiotti:2016qnr,Alford:2017rxf,Most:2022yhe}, and accretion disks around black holes \cite{Foucart:2017axc}. One of the main open questions in the field is understanding whether (and how) such fluid dynamic description can also be applied to exotic systems that can be far from equilibrium, such as the quark-gluon plasma formed in high-multiplicity proton-nucleus and proton-proton collisions \cite{CMS:2015yux,ATLAS:2015hzw,CMS:2016fnw,Weller:2017tsr,PHENIX:2018lia,ALICE:2019zfl,Noronha:2024dtq,Grosse-Oetringhaus:2024bwr} and weakly collisional astrophysical plasmas \cite{Chandra:2015iza,Foucart:2017axc,EventHorizonTelescope:2022wkp,Cordeiro:2023ljz}. Despite extensive research into the formulation of relativistic hydrodynamics, the underlying physics and mathematics still pose significant challenges. In particular, especially when considering applications to physical systems, it is crucial to understand the validity of hydrodynamic theories in regimes that extend beyond the linear regime around equilibrium (for a field theory perspective concerning causality and diffusion, see \cite{abbasi2025}).

The first formulations of relativistic hydrodynamics can be traced back to the first-order formulations of Eckart \cite{EckartViscous} and Landau and Lifshitz \cite{LandauLifshitzFluids}. However, both Eckart and Landau theories are known to be acausal and unstable, making them unsuitable for numerical simulations of viscous fluids in the relativistic regime \cite{Hiscock_Lindblom_stability_1983,Hiscock_Lindblom_instability_1985,Hiscock_Lindblom_acausality_1987}. Alternative formulations such as Israel-Stewart (IS) theory \cite{MIS-2,MIS-6} and, more recently, Bemfica-Disconzi-Noronha-Kovtun (BDNK) theory \cite{Bemfica:2017wps,Kovtun:2019hdm,Bemfica:2019knx,Hoult:2020eho,Bemfica:2020zjp,Abboud:2023hos}, have been developed to address these limitations. 

In BDNK theory, different than Eckart and Landau and Lifshitz, the constitutive relations contain all the possible first-order terms in derivatives of the hydrodynamic variables of ideal fluid dynamics that are compatible with the symmetries. This defines first-order hydrodynamics in a general hydrodynamic frame. One can then prove \cite{Bemfica:2017wps,Kovtun:2019hdm,Bemfica:2019knx,Hoult:2020eho,Bemfica:2020zjp,Abboud:2023hos} that there are hydrodynamic frame choices (which are different than the choices made by Eckart and Landau and Lifshitz) where causality and stability hold.  

The problems related to the acausality and instability of Eckart and Landau theory are overcome by IS hydrodynamics by assuming that dissipative currents represent new dynamical degrees of freedom that obey nonlinear relaxation equations that can be derived in various ways, see Refs.\ \cite{MIS-2,MIS-6,Baier:2007ix,Baier:2007ix,Denicol:2012cn,Noronha:2021syv}. These equations, together with the conservation laws, describe how dissipative quantities relax to their relativistic Navier-Stokes limits over some specific timescales (the relaxation times).

Israel-Stewart-like formulations are widely used in nuclear theory applications in the context of heavy-ion collisions, see e.g. \cite{Romatschke:2017ejr}. However, despite intense numerical investigations through the last decade, it is important to note that most of what is understood about IS-like theories stems from the so-called linear regime, where only small perturbations around equilibrium are considered, see e.g. \cite{Brito:2020nou,deBrito:2025jaz}. In fact, in the full nonlinear regime, it is not known if theories of IS-type\footnote{In this work, any formulation of relativistic viscous fluid dynamics in which dissipative currents obey additional equations of motion, such as the original IS-theory \cite{MIS-2,MIS-6}, resummed BRSSS \cite{Baier:2007ix}, and DNMR theory \cite{Denicol:2012cn}, is called of Israel-Stewart type.} generally admit solutions that are unique given suitable initial data (this is part of the fundamental question concerning the local well-posedness of the Cauchy problem in such theories). 

There is limited knowledge regarding the \emph{nonlinear} properties of IS-like theories, with only a few general results established in the literature. In the context of heavy-ion collisions, nonlinear causality was investigated in highly symmetric situations in \cite{Floerchinger_2018}. Without simplifying assumptions on geometry, general bounds defining causality and local well-posedness in the nonlinear regime of IS-like theories have been derived for the case of bulk viscosity in \cite{Bemfica:2019cop}, where the theory can be proved to be symmetric hyperbolic. For such theories with bulk viscosity, it was proved in \cite{Disconzi:2020ijk} that there exists a class of smooth initial data for
which solutions develop a singularity in finite time
or become acausal. Separate sets of necessary and sufficient conditions that ensure nonlinear causality in the case involving both shear and bulk viscosity (at vanishing chemical potential) were obtained in \cite{Bemfica:2020xym}. These conditions have been instrumental in clarifying the regime of validity of IS-like theories in simulations of the quark-gluon plasma, see \cite{Plumberg:2021bme,Chiu:2021muk,Krupczak:2023jpa,Domingues:2024pom}. General nonlinear results concerning causality and well-posedness have also been obtained in the context of a particular class of magnetohydrodynamic theories with shear viscosity \cite{Cordeiro:2023ljz} (proven to be strongly hyperbolic), which can be generalized in a straightforward manner to include bulk viscosity as well.

It is essential to note that while \emph{linear} causality and stability analyses of IS-like theories, e.g. \cite{Hiscock_Lindblom_stability_1983}, place meaningful constraints on transport coefficients and the equation of state, a full nonlinear analysis in IS-like theories also places constraints on the dissipative currents themselves (affecting, thus, the choice of initial data and the subsequent evolution), see \cite{Bemfica:2019cop,Floerchinger_2018,Bemfica:2020xym,Cordeiro:2023ljz}. This can restrict or even rule out relevant physical processes, such as plasma instabilities in the context of magnetohydrodynamics \cite{Cordeiro:2023ljz}, which are inaccessible via linear analyses.

In this paper, we present the first full analysis of nonlinear causality for IS theory with a conserved vector (baryon). In principle, one could extend this discussion to other conserved currents, such as \cite{Greif:2017byw,Almaalol:2022pjc,Capellino_2022,Capellino_2023}). For the sake of simplicity, we consider the case where the equations of motion for the dissipative currents are obtained from imposing that a suitable nonequilibrium entropy current has non-negative entropy production \cite{MIS-6} (though our results can be generalized for other sets of equations of motion as well). We study two specific cases: the Landau hydrodynamic case, which considers only number diffusion, and the Eckart frame, which focuses solely on energy diffusion. We provide causality constraints in $D = 3 + 1$ dimensions that directly limit the size of both equilibrium and dissipative dynamics without any assumptions on the spacetime geometry or equation of state of the system with diffusion in both frames. Furthermore, we show how a redefinition of the dynamic fields between Landau and Eckart results in a region of causality that differs both structurally and physically. In the Landau frame, we show that these constraints entirely specify the region of nonlinear causality, whereas the Eckart frame's bounds are impeded by an order $5$ polynomial which cannot be solved analytically in general, meaning that a complete specification of the region is not possible without resorting to numerical investigations.



For both Landau and Eckart frames, we show that our constraints recover results from previous nonlinear analyses done in $1 + 1$ dimensions for a particular ultrarelativistic equation of state \cite{HISCOCK1989125,Hiscock_Lindblom_pathologies_1988}. In the Landau frame, we show that the $D = 1 + 1$ constraints are necessary \emph{and} sufficient for the $D = 3 + 1$ nonlinear causality conditions for an ultrarelativistic ideal gas. Within this framework, there exists a region of nonlinear causality that allows for the (causally allowed) spacelike propagation of baryon current which was previously obtained in \cite{HISCOCK1989125}. Following \cite{Hiscock_Lindblom_pathologies_1988} in the Eckart frame, we show that an ultrarelativistic ideal gas in $D = 1 + 1$ dimensions provides necessary but \emph{not} sufficient conditions for our causality conditions in $D = 3 + 1$. In addition, an analogous calculation in the Eckart frame suggests that nonlinear causality ensures that the ultrarelativistic ideal gas never violates the dominant energy condition---a contrast to the spacelike baryon current in the Landau frame. In other words, this difference shows that constraints coming from nonlinear causality depend on the choice of hydrodynamic frame. Finally, we derive explicit expressions for the linearized causality bounds in both hydrodynamic frames, and show that the linear conditions fail to capture the new physical constraints on energy and
number diffusion that are obtained through a nonlinear causality analysis. \\

This paper is organized as follows. In Sec.~\ref{sec:HydrodynamicFrames}, we review the Landau and Eckart hydrodynamic frames along with the relevant degrees of freedom prescribed by them. In Sec.~\ref{sec:LandauFrame}, we derive the general necessary and sufficient nonlinear causality bounds in $D = 3 + 1$ under the Landau frame prescription and relate them to a calculation done originally in \cite{HISCOCK1989125} for an ultrarelativistic ideal gas in $D = 1 + 1$ dimensions. We also provide explicit expressions for the linear causality conditions provided implicitly in \cite{Olson:1989ey} for the general Landau frame theory with all dissipative fluxes. In Sec.~\ref{sec:EckartFrame}, we provide analogous yet distinct nonlinear causality constraints for the Eckart frame and show that these conditions imply the dominant energy condition is fulfilled for the ultrarelativistic gas in $D = 1 + 1$ dimensions considered previously in \cite{Hiscock_Lindblom_pathologies_1988}. Finally, we explicitly derive analogous Eckart frame linear causality conditions provided implicitly in \cite{Hiscock_Lindblom_stability_1983}. We finish with our conclusions in Sec.~\ref{sec:Conclusion}. Appendices are added to give further details about the mathematical proofs presented in the main text.

\emph{Notation}: We use natural units $\hbar=c=k_B=1$ and our (arbitrary) spacetime metric $g_{\mu\nu}$ has a mostly plus signature. 

\section{Hydrodynamic Frames}\label{sec:HydrodynamicFrames}

Let us consider relativistic fluids described by an energy-momentum tensor $T^{\mu\nu}$ and a baryon current $J^\mu$ \cite{Rezzolla_Zanotti_book}. The resulting relativistic hydrodynamic equations of motion are thus given by the conservation laws
\begin{equation}
    \nabla_\alpha T^{\alpha\mu} = 0, \qquad \qquad \nabla_\alpha J^\alpha =0.
\end{equation}
These equations govern the evolution of the degrees of freedom, such as the fluid four-velocity $u^\mu$ (normalized such $u^\mu u_\mu = -1$), equilibrium energy density $\varepsilon$, pressure $P$, and charge density $n$. Note that only two among $\varepsilon$, $P$, and $n$ are treated as independent variables as they are related through the equilibrium equation of state, so $P = P(\varepsilon,n)$. In this paper, we focus on the relativistic Israel-Stewart theories \cite{MIS-2,MIS-6} that describe the dissipation processes associated with a vector dissipative quantity. The interpretation of the vector quantity in question depends on the choice of hydrodynamic frame \cite{MIS-6}: in the Landau frame this corresponds to particle diffusion $\mathcal{J}^\mu$, while in the Eckart frame one encounters energy diffusion $q^\mu$.


In the Landau frame, the baryon number diffusion current $\mathcal{J}^\mu$ (with $\mathcal{J}^\alpha u_\alpha = 0$) is added to $J^\mu$, though $u^\mu$ is an eigenvector of the energy-momentum tensor (as it was the case in ideal hydrodynamics). Thus, the Landau frame is also known as the energy-frame \cite{LandauLifshitzFluids}. The explicit expressions for the energy-momentum tensor and current are given by
\bml
\label{eq:LandauConservedCharges}
\bea
T^{\mu\nu} &=& \varepsilon u^\mu u^\nu + P\Delta^{\mu\nu},\\
J^\mu &=& nu^\mu + \mathcal{J}^\mu,
\eea
\eml
where we defined $\Delta_{\mu\nu} = g_{\mu\nu} + u_\mu u_\nu$ as the projection tensor orthogonal to $u^\mu$. 

On the other hand, the Eckart frame is known as the particle frame, as it instead defines the four-velocity fluid so that it remains parallel to the full baryon current $J^\mu$ throughout the evolution \cite{EckartViscous}. In this case, the energy-momentum tensor and current are given by
\bml
\label{eq:EckartConservedCharges}
\bea
T^{\mu\nu} &=& \varepsilon u^\mu u^\nu + P\Delta^{\mu\nu} + q^\mu u^\nu + u^\mu q^\nu,\\
J^\mu &=& nu^\mu.
\eea
\eml
with dissipation governed by the energy/heat diffusion vector $q^\mu$ (satisfying $q^\alpha u_\alpha = 0$). For the sake of clarity, we emphasize that in this work we will only focus on the consequences of energy/particle diffusion, so no effects from shear and/or bulk viscosity are taken into account. We note that nonlinear causality constraints in the absence of diffusion but with shear and bulk viscosity (at zero chemical potential) have been previously derived in Ref.~\cite{Bemfica:2020xym}. Our present work complements that analysis by focusing on the diffusive sector, which introduces distinct structural features in the characteristic determinant, particularly in how hydrodynamic frame choices affect causality bounds.

The Landau and Eckart frame four-velocity [which here we will denote with sub/superscripts $(L)$ and $(E)$] are related up to first order in dissipative currents \cite{ISRAEL1976310} via the transformation
\be
u_{(L)}^\mu = u_{(E)}^\mu + \frac{q^\mu}{\varepsilon + P} + \mathcal{O}(q^2).
\ee
It is important to stress that the notion of a hydrodynamic frame should not be interpreted solely as a choice of computational convenience, especially in the case of first-order theories. In fact, one can show that some frame choices can lead to acausality and instability \cite{Hiscock_Lindblom_acausality_1987,Hiscock_Lindblom_instability_1985}, whereas others do not \cite{Bemfica:2017wps,Kovtun:2019hdm,Bemfica:2019knx,Hoult:2020eho,Bemfica:2020zjp,Abboud:2023hos}. One should therefore think of fixing different hydrodynamic frames as fixing different effective descriptions of a theory, resulting in a different physical understanding of the dynamic quantities in question. In the following sections, we discuss how the choice of hydrodynamic frame in second-order hydrodynamics impacts the nonlinear causal evolution of the initial data subject to the Landau and Eckart frame in the presence of number and energy diffusion, respectively.

\section{Causality in the Landau Frame}\label{sec:LandauFrame}

In the Landau prescription, the energy ($u_\beta\nabla_\alpha T^{\alpha \beta} = 0$), momentum ($\tensor{\Delta}{^\mu_\beta}\nabla_\alpha T^{\alpha\beta} = 0$), and baryon conservation ($\nabla_\alpha J^\alpha = 0$) equations may be cast as
\bml
\label{Eq_Landau}
\bea
\label{eq:Lmomentum}
0 &=& \tensor{g}{^\mu_\nu}u^\alpha\nabla_\alpha u^\nu + \frac{\Delta^{\alpha \mu}P_\varepsilon}{\varepsilon +P}\nabla_\alpha \varepsilon + \frac{\Delta^{\alpha \mu}P_n}{\varepsilon +P}\nabla_\alpha n,\\
\label{eq:Lenergy}
0 &=&  (\varepsilon +P)\tensor{g}{^\alpha_\nu}\nabla_\alpha u^\nu + u^\alpha\nabla_\alpha \varepsilon,\\
\label{eq:Lbaryon}
0 &=& n\tensor{g}{^\alpha_\nu}\nabla_\alpha u^\nu + u^\alpha\nabla_\alpha n + \tensor{g}{^\alpha_\nu}\nabla_\alpha\mathcal{J}^\nu,
\eea
\eml
where we have assumed the normalization condition $u^\alpha\nabla^\mu u_\alpha = 0$. Here, we have implicitly assumed that all scalars $C$ can be parametrized (in a smooth, invertible sense) in terms of $\varepsilon$ and $n$ such that $C \equiv C(\varepsilon,n)$. We then use the convenient shorthand $C_\varepsilon \equiv \left(\frac{\partial C}{\partial\varepsilon}\right)_n$ and $C_n \equiv \left(\frac{\partial C}{\partial n}\right)_\varepsilon$ for any such scalar. In the Israel-Stewart formulation \cite{ISRAEL1976310,ISRAEL1979341}, $\mathcal{J}^\mu$ is promoted to an independent degree of freedom governed by a \emph{relaxation equation} of the form
\be
\label{eq:Ldiffusion}
\frac{\tau_{\mathcal{J}}}{\kappa T^2}u^\alpha\left(\nabla_\alpha \mathcal{J}^\mu - u^\mu \mathcal{J}^\nu \nabla_\alpha u_\nu\right)+ \frac{\mathcal{J}^\mu}{\kappa T^2} = -\Delta^{\mu\alpha}\nabla_\alpha\Theta - \frac{\Delta^{\mu\beta}}{2}\mathcal{J}_\beta\nabla_\alpha\left(u^\alpha \frac{\tau_{\mathcal{J}}}{\kappa T^2}\right),
\ee
where $\tau_\mathcal{J}$ is the relaxation time, $\kappa$ is the diffusion coefficient, and $\Theta = \mu/T$ is the ratio between chemical potential and temperature.
Israel and Stewart's formulation follows an entropy current $\mathcal{S}^\mu$ argument that expands up to second-order in dissipative (e.g. nonequilibrium) fluxes $\mathcal{J}^\mu$ such that
\be
\label{eq:Lentropycurrent}
T\mathcal{S}^\mu = \left(Ts - \frac{\tau_{\mathcal{J}}}{\kappa T}\frac{\mathcal{J}^2}{2}\right)u^\mu - \mu\mathcal{J}^\mu,
\ee
where $\mu,s$ and $T$ are the chemical potential, entropy density, and temperature respectively. Imposing the second law of thermodynamics for this entropy current in the form
\be
\nabla_\alpha\mathcal{S}^\alpha = \frac{\tau_{\mathcal{J}}}{\kappa T}\mathcal{J}^2 \geq 0;\quad \kappa > 0,\:\tau_{\mathcal{J}}\geq 0,
\ee
leads to Eq.~\eqref{eq:Ldiffusion}, where $\mathcal{J}^2\equiv \mathcal{J}^\alpha\mathcal{J}_\alpha$. 

\subsection{The characteristic determinant}

The equations of motion in Eq.~\eqref{eq:Lenergy}--\eqref{eq:Lbaryon}, and \eqref{eq:Ldiffusion} can be uploaded into a matrix equation of the form
\be
\label{eq:LPDE}
\mathbb{A}_L^\alpha\partial_\alpha\Umb + \mathbb{B}_L\Umb = \boldsymbol{0}
\ee
where $\boldsymbol{0}\in\mathbb{R}^{10}$ is the zero vector, $\Umb\in\mathbb{R}^{10}$ is the vector of dynamic quantities given by
\be
\label{eq:Ldynamics}
\Umb \equiv
\begin{pmatrix}
u^\nu\\
\mathcal{J}^\nu\\
\varepsilon\\
n
\end{pmatrix}
\ee
with raised indices denoting representative elements of column vectors, and lowered corresponding to row vectors. Here, $\mathbb{A}_L^\mu$ and $\mathbb{B}_L$ are $10\times 10$ real matrices whose entries are nonlinear expressions involving $\Umb$ but not its derivatives. In this sense, the system is a quasilinear partial differential equation, as its highest order derivative (first) terms appear linearly, with all lower order terms allowed to be nonlinear \cite{ChoquetBruhatGRBook}. The matrix $\mathbb{B}_L$, along with a suitable (i.e. locally well-posed) Cauchy problem \cite{ChoquetBruhatGRBook} fixes a particular solution, and further mathematical/physical properties of the system can be garnered from the highest order derivative term, known as the \emph{principal part} $\mathbb{A}^\alpha\phi_\alpha$, where we write $\partial_\mu\Umb\rightarrow \phi_\mu$. The characteristics of the system are defined by the solutions of the characteristic equation $\det(\mathbb{A}^\alpha\phi_\alpha) =0$ \cite{ChoquetBruhatGRBook}. The principal part can be expressed in matrix form\footnote{For clarity, we note that $0^\mu_\nu$ denotes a 4 by 4 matrix in which all elements are zero.} as
\be
\label{eq:Lprinciplepart}
\mathbb{A}_L^\alpha\phi_\alpha=
\begin{pmatrix}
x\tensor{g}{^\mu_\nu}& \tensor{0}{^\mu_\nu}& \frac{P_\varepsilon}{\varepsilon + P}v^\mu & \frac{P_n}{\varepsilon + P} v^\mu\\
\frac{\mathcal{J}^\mu\phi_\nu}{2} - x u^\mu\mathcal{J}_\nu& x\tensor{g}{^\mu_\nu}& \frac{x\Omega_{\mathcal{J},\varepsilon}}{2\Omega_{\mathcal{J}}}\mathcal{J}^\mu + \frac{\Theta_\varepsilon}{\Omega_{\mathcal{J}}} v^\mu& \frac{x\Omega_{\mathcal{J},n}}{2\Omega_{\mathcal{J}}}\mathcal{J}^\mu + \frac{\Theta_n}{\Omega_{\mathcal{J}}} v^\mu\\
(\varepsilon + P)\phi_\nu& 0_\nu& x& 0\\
n\phi_\nu& \phi_\nu& 0& x
\end{pmatrix}.
\ee
Here, $\Omega_{\mathcal{J}}\equiv \tau_{\mathcal{J}}/\kappa T^2$, and we have defined the following scalar products:
\be
x = u^\alpha \phi_\alpha,\quad y_{\mathcal{J}} = \mathcal{J}^\alpha \phi_\alpha,\quad v^\mu = \Delta^{\mu\alpha}\phi_\alpha.
\ee
Since $v^\mu$ and $\mathcal{J}^\mu$ lie in the subspace orthogonal to $u^\mu$, note that $\Delta_{\mu\nu}$ defines an inner product between them, meaning that there must exist some $\psi\in[-1,1]$ such that $y_{\mathcal{J}} = \mathcal{J}^\alpha\phi_\alpha = \mathcal{J}^\alpha v_\alpha = \psi\mathcal{J}v$ where we analogously write $v^2\equiv v^\alpha v_\alpha$. The determinant of the principal part can then be found to be
\be
\label{eq:LandauDet}
\det(\mathbb{A}_L^\alpha\phi_\alpha) = v^{10}\hat{x}^6P_4^{(L)}(\hat{x};\psi),
\ee
where $\hat{x}\equiv x/v$ and 
\be
P_4^{(L)}(\hat{x};\psi) = \sum_{a = 0}^4\mathcal{A}_a^{(L)}(\psi) \hat{x}^a,
\ee
where $\mathcal{A}_4^{(L)} = 1$ and 
\bml
\label{eq:LCoeffs}
\bea
\mathcal{A}_3^{(L)} &=& -\psi\mathcal{J}\left(\frac{P_n}{\varepsilon + P} + \frac{\Omega_{\mathcal{J},n}}{2\Omega_{\mathcal{J}}}\right),\\
\mathcal{A}_2^{(L)} &=& -c_s^2 - \frac{\Theta_n}{\Omega_{\mathcal{J}}},\\
\mathcal{A}_1^{(L)} &=& \frac{\psi\mathcal{J}}{2} \left(\frac{P_n}{\varepsilon + P} + \frac{\Omega_{\mathcal{J},n}}{\Omega_{\mathcal{J}} } c_s^2-\frac{\Omega_{\mathcal{J},\varepsilon|\bar{s}}}{\Omega_{\mathcal{J}}} P_n \right),\\
\mathcal{A}_0^{(L)} &=& \frac{c_s^2 \Theta_n-\Theta_{\varepsilon|\bar{s}} P_n}{\Omega_{\mathcal{J}}}.
\eea
\eml
Let $\bar{s}\equiv S/N$ be the specific entropy (in contrast to the entropy density $s = S/V$). Above, we applied the convenient substitutions (note that $c_s$ is the speed of sound at constant entropy per particle)
\bml
\bea
\Omega_{\mathcal{J},\varepsilon|\bar{s}} &\equiv & \left(\frac{\partial\Omega_{\mathcal{J}}}{\partial\varepsilon}\right)_{\bar{s}} = \Omega_{\mathcal{J},\varepsilon} + \frac{n\Omega_{\mathcal{J},n}}{\varepsilon + P},\\
c_s^2 &\equiv & \left(\frac{\partial P}{\partial\varepsilon}\right)_{\bar{s}} = P_\varepsilon + \frac{nP_n}{\varepsilon + P},\\
\Theta_{\varepsilon|\bar{s}} &\equiv & \left(\frac{\partial\Theta}{\partial\varepsilon}\right)_{\bar{s}} = \Theta_\varepsilon + \frac{n\Theta_n}{\varepsilon + P}.
\eea
\eml

\subsection{Nonlinear causality in the Landau frame}

Broadly speaking, causality is the fundamental property that enforces that information cannot propagate faster than the speed of light. In terms of PDEs, causality is enforced by requiring that the characteristic vectors are always nontimelike. These characteristics may be interpreted as the normal vectors $\phi^\mu$ to the solution surface: i.e. $\partial_\mu \Umb\leftrightarrow \phi_\mu$ where $\Umb$ is the solution vector, meaning that the solution surface is then timelike. This notion may be formulated by the following definition \cite{Bemfica:2020zjp}: a system of the form $(\mathbb{A}^\alpha\partial_\alpha + \mathbb{B})\Umb = \boldsymbol{0}$ evolves \emph{causally} if and only if for any characteristic vector $\phi^\mu$ defined by the characteristic equation $\det(\mathbb{A}^\alpha\phi_\alpha) = 0$ 
\bml
\label{eq:causalitydef}
\bea
\textrm{(CI)}&&\textrm{ the roots $\phi_0 \equiv \phi_0(\phi_j)$ of the characteristic equation exist in $\mathbb{R}$ and}\\
\textrm{(CII)}&&\textrm{ $\phi^\alpha\phi_\alpha \geq 0$.}
\eea
\eml
Here an important digression is necessary. Consider a solution $(\varepsilon, n, u^\nu, \mathcal{J}^\nu)$ of the original equations $\nabla_\nu T^{\mu\nu} = \nabla_\mu J^\mu = 0$, together with the relaxation equation for $\mathcal{J}^\mu$, defined on a globally hyperbolic spacetime $M$ \cite{ChoquetBruhatGRBook}. In the original system, $u^0$ and $\mathcal{J}^0$ are nondynamical constrained variables satisfying $u^\mu u_\mu + 1 = u_\nu \mathcal{J}^\nu = 0$, which the given solution respects. Remarkably, the solution $(\varepsilon, n, u^\nu, \mathcal{J}^\nu)$ also satisfies the extended system of equations given in \eqref{Eq_Landau} and \eqref{eq:Ldiffusion}, despite $u^0$ and $\mathcal{J}^0$ now being treated as dynamical variables. 
It is crucial to emphasize that the extended system serves exclusively as a \emph{mathematical tool} for establishing causality of solutions to the original equations. While the extended system admits a broader class of solutions, only those satisfying the original constraints $u^\mu u_\mu + 1 = u_\nu \mathcal{J}^\nu = 0$ are physically relevant. The utility of this extension lies in the fact that
(i) the causality conditions (\ref{eq:causalitydef}) for the extended system are easier to formulate and verify, (ii) any causal behavior proved for the extended system automatically transfers to the constrained (original) solutions, and (iii) the additional, unphysical solutions of the extended system are irrelevant for our purposes---they merely represent a mathematical convenience in the proof structure. Thus, while working with the extended system, we temporarily enlarge our variable space purely as a technical device. The physical content remains entirely contained in the original system's solution space, and the extended system's additional degrees of freedom have no bearing on the actual physical theory once the causality argument is complete.

That said, causality is then established in the following sense. Let $\Sigma$ be a Cauchy surface \cite{ChoquetBruhatGRBook}. The causality conditions \eqref{eq:causalitydef} guarantee that for any point $p \in M$ in the future of $\Sigma$, the field values $(\varepsilon(p), n(p), u^\nu(p), \mathcal{J}^\nu(p))$ depend solely on the initial data $(\varepsilon, n, u^\nu, \mathcal{J}^\nu)$ restricted to $\Sigma \cap J^-(p)$, where $J^-(p)$ denotes the causal past of $p$. Consequently, establishing causality for the extended system ensures causality for solutions of the original system, upon which the constraints $u^\mu u_\mu + 1 = u_\nu \mathcal{J}^\nu = 0$ are properly maintained under evolution.

Guaranteeing the reality of the roots of the characteristic equation is a straightforward result from algebra \cite{Rees_1922}. We shall list these conditions below in the case of the Landau frame:
\begin{proposition}\label{prop:Lreal}
Let $\psi\in[-1,1]$, $\Delta_L\equiv \Delta[P_4^{(L)}]$ be the discriminant of $P_4^{(L)}$ and define the following quantities:
\bml
\bea
\widetilde{A}_{2}^{(L)}(\psi^2) &=& -\frac{3\psi^2 \mathcal{J}^2}{8} \left(\frac{P_n}{\varepsilon + P}+\frac{\Omega_{\mathcal{J},n}}{2\Omega_{\mathcal{J}}}\right)^2-\frac{\Theta_n}{\Omega_{\mathcal{J}}}-c_s^2\\
\delta_0^{(L)}(\psi^2) &=& \frac{3}{2} \psi^2 \mathcal{J}^2 \left(\frac{P_n}{\varepsilon + P}+\frac{1}{2}\frac{\Omega_{\mathcal{J},n}}{\Omega_{\mathcal{J}}} \right) \left(\frac{P_n}{\varepsilon + P}+ \frac{\Omega_{\mathcal{J},n}}{\Omega_{\mathcal{J}}} c_s^2-\frac{\Omega_{\mathcal{J},\varepsilon|\bar{s}}}{\Omega_{\mathcal{J}}}P_n\right)\notag\\
&&\quad + c_s^4+2 \left(7 c_s^2 \frac{\Theta_n}{\Omega_{\mathcal{J}}}-6 \frac{\Theta_{\varepsilon|\bar{s}}}{\Omega_{\mathcal{J}}}P_n\right)+ \frac{\Theta_n^2}{\Omega_{\mathcal{J}}^2}\\
\delta_1^{(L)}(\psi^2) &=&-\psi^2 \mathcal{J}^2 \left(\frac{P_n}{\varepsilon + P} + \frac{\Omega_{\mathcal{J},n}}{2\Omega_{\mathcal{J}}} \right) \left[\left(c_s^2-\frac{1}{2}\right) \frac{P_n}{\varepsilon + P}+ P_n\frac{\Omega_{\mathcal{J},\varepsilon|\bar{s}}}{2\Omega_{\mathcal{J}}} + \frac{P_n}{\varepsilon + P}\frac{\Theta_n}{\Omega_{\mathcal{J}}}+\frac{\Omega_{\mathcal{J},n}\,\Theta_n}{2\Omega_{\mathcal{J}}^2}\right] \notag\\
&&\quad -\left(\left(\frac{\Theta_n}{\Omega_{\mathcal{J}}}- c_s^2\right)^2+4 \frac{\Theta_{\varepsilon|\bar{s}}}{\Omega_{\mathcal{J}}} P_n\right)-\frac{3 \psi^4\mathcal{J}^4 }{16}\left( \frac{P_n}{\varepsilon + P}+\frac{\Omega_{\mathcal{J},n}}{2\Omega_{\mathcal{J}}}\right)^4.
\eea
\eml
The roots of the characteristic polynomial defined via the characteristic equation $\det(\mathbb{A}^\alpha_L\phi_\alpha) = 0$ are real if and only if one of the following conditions holds. Furthermore, these conditions provide information of the multiplicity of the roots as provided below.
\begin{enumerate}
    \item If $\Delta_L > 0$ and $\widetilde{A}_{2}^{(L)},\delta_1^{(L)} < 0$, then all roots are real and distinct.

    \item If $\Delta_L = 0$ and one of the following conditions hold, then all roots are real, with some nondistinct roots: 
    
    \begin{itemize}
        \item If $\widetilde{A}_{2}^{(L)},\delta_1^{(L)} < 0$ and $\delta_0^{(L)}\neq 0$, then there is a real double root, and two real simple roots (3 distinct).

        \item If $\delta_0^{(L)} = 0$ and $\delta_1^{(L)} \neq 0$, there is a triple root and one simple root which are all real (2 distinct).

        \item If $\delta_1^{(L)} = 0$ and $\widetilde{A}_{2}^{(L)} < 0$, there are two double real roots (2 distinct).

        \item If $\delta_0^{(L)} = \delta_1^{(L)} = 0$ there is one real root with multiplicity 4 (1 distinct).
    \end{itemize}
\end{enumerate}
\begin{proof}
See \cite{Rees_1922} for a detailed discussion of necessary and sufficient conditions for existence of quartic roots.
\end{proof}
\end{proposition}
\begin{theorem}\label{thm:LandauCausality}
Suppose that the roots of the characteristic equation $\det(\mathbb{A}_L^\alpha\phi_\alpha) = 0$ are real, that is, they satisfy one of the conditions in proposition~\ref{prop:Lreal} for all $\psi\in[-1,1]$. Then the system $(\mathbb{A}_L^\alpha\partial_\alpha + \mathbb{B}_L)\Umb = \boldsymbol{0}$ is causal if and only if the following bounds are satisfied:
\bml
\bea
\frac{\mathcal{J}}{4} \left|\frac{P_n}{\varepsilon + P} + \frac{\Omega_{\mathcal{J},n}}{2\Omega_{\mathcal{J}}}\right| &\leq & \min\left\{1,\frac{1}{2}\left(1-\frac{c_s^2}{6}\right) - \frac{\Theta_n}{12\Omega_{\mathcal{J}}}\right\},\\
\frac{\mathcal{J}}{4} \left|\frac{\Omega_{\mathcal{J},n}}{\Omega_{\mathcal{J}}}\left(3-c_s^2\right) + \frac{5P_n}{\varepsilon + P} + \frac{\Omega_{\mathcal{J},\varepsilon|\bar{s}}}{\Omega_{\mathcal{J}}}P_n\right| &\leq & \left(1-c_s^2\right) + \left(1-\frac{\Theta_n}{\Omega_{\mathcal{J}}}\right),\\
\frac{\mathcal{J}}{2} \left|\frac{\Omega_{\mathcal{J},n}}{\Omega_{\mathcal{J}}}\left(1-c_s^2\right) + \frac{P_n}{\varepsilon + P} + \frac{\Omega_{\mathcal{J},\varepsilon|\bar{s}}}{\Omega_{\mathcal{J}}}P_n\right| &\leq & (1-c_s^2)\left(1 -\frac{\Theta_n}{\Omega_{\mathcal{J}}}\right) -\frac{P_n}{\Omega_{\mathcal{J}}}\Theta_{\varepsilon|\bar{s}}.
\eea
\eml
\begin{proof}
The above conditions are necessary and sufficient for bounding the characteristic wave speeds (which are the roots of the characteristic equation) between $\pm 1$. These conditions are found by imposing (CII) in Eq.~\eqref{eq:causalitydef} from our definition of causality and then showing that the inequalities satisfied by the coefficients $\mathcal{A}_a^{(L)}$ for $a = 0,1,2,3$ are equivalent conditions naturally arising from the polynomial in $\hat{x}$ itself. The interested reader may find more information in  Appendix~\ref{App:LCausality}.
\end{proof}
\end{theorem}

The results of this section illustrate the general idea that, in the nonlinear regime, causality places constraints not only on the transport coefficients and the equation of state, but also on the magnitude of the dissipative currents themselves. This should be contrasted to the standard results obtained in the linearized regime, see e.g. \cite{Hiscock_Lindblom_stability_1983,Olson:1989ey}, which do not depend on $\mathcal{J}$. 

\subsection{Ultrarelativistic ideal gas}\label{subsec:URLandau}

Nonlinear causality of Israel-Stewart with number diffusion has previously been analyzed for a single spatial dimension and ultrarelativistic ideal gas equation of state in Ref.\ \cite{HISCOCK1989125}. In this section we focus on this particular ultrarelativistic ideal gas equation of state and show that a particular choice of orientation of $\mathcal{J}^\mu$ recovers the $D = 1 + 1$ characteristic determinant in \cite{HISCOCK1989125} and the $D = 1 + 1$ constraints provide a necessary \emph{and sufficient} constraint for our $D = 3 + 1$ nonlinear causality bounds for the ultrarelativistic ideal gas equation of state. Furthermore, we recreate a brief calculation done in \cite{HISCOCK1989125} showing that the total baryon number current $J^\mu$ is allowed to become spacelike even though nonlinear causality remains upheld. Since the bounds for this $1 + 1$ case are necessary and sufficient for the $3 + 1$  case, this result implies the somewhat surprising result that having a spacelike baryon current is not forbidden by nonlinear causality in this case.

For an ultrarelativistic ideal gas of baryons, with equation of state $P(\varepsilon) = \varepsilon/3 = nT(\varepsilon,n)$, the nontrivial quartic portion of Eq.~\eqref{eq:LandauDet} decouples into a product of quadratics
\be
\label{eq:URLandauDet}
P_4^{(L)}(\hat{x},\psi)\Bigg|_{P(\varepsilon),T(\varepsilon,n)} = \left(\hat{x}^2 - \frac{1}{3}\right)\left(\hat{x}^2-\frac{\Omega_{\mathcal{J},n}}{2 \Omega_{\mathcal{J}}}\psi\mathcal{J}\hat{x}-\frac{4}{n\Omega_{\mathcal{J}}}\right).
\ee
The roots corresponding to $\hat{x} = \pm 1/\sqrt{3}$ are causal as they are real and also less than unity in magnitude. For the nontrivial quadratic, the reality of the roots is guaranteed by the non-negativity of the discriminant $\Delta_2$ and the supposition that the roots are bounded by unity in magnitude. The following are then necessary and sufficient causality conditions for the ultrarelativistic case
\bml
\bea
\Delta_2(\psi) \equiv\frac{16}{\Omega_{\mathcal{J}} n}+\frac{n^2\Omega_{\mathcal{J},n}^2}{4 \Omega_{\mathcal{J}}^2}\frac{\psi^2\mathcal{J}^2}{n^2} &\geq & 0,\\
\frac{1}{2}\left|\frac{n\Omega_{\mathcal{J},n}}{2\Omega_{\mathcal{J}}}\psi\frac{\mathcal{J}}{n}\pm \sqrt{\Delta_2(\psi)}\right| &\leq & 1.
\eea
\eml
The $1 + 1$ dimensional case mentioned above is equivalent to setting $\psi = 1$ such that the number diffusion remains parallel to the component of the characteristic vector orthogonal to the fluid trajectory: $y = \mathcal{J}^\alpha v_\alpha = \mathcal{J}v$. Since $\psi\in[-1,1]$, the triangle inequality $|a + b|\leq |a| + |b|$ for $a,b\in\mathbb{R}$ paired with $|\psi| = 1$ provides the most stringent constraint on the system of inequalities. Furthermore, at linear order in fluctuations from equilibrium, the Israel-Stewart transport coefficient $\Omega_{\mathcal{J}}$ is unstable for negative values \cite{Hiscock_Lindblom_stability_1983}, meaning that we will assume henceforth that $\Omega_{\mathcal{J}}\geq 0$. This constraint guarantees that $\Delta_2$ is always non-negative, and also reduces the remaining nonlinear causality bounds to the single constraint
\be
\label{eq:UR1Dcausality}
\frac{1}{2}\left|\frac{\mathcal{J}}{n}\right|\leq \dfrac{n\Omega_{\mathcal{J}} - 4}{n^2\left|\Omega_{\mathcal{J},n}\right|}.
\ee
In summary, the $1 + 1$ dimensional nonlinear causality bound \eqref{eq:UR1Dcausality} provides the most stringent bound, and consequently, a necessary and sufficient constraint for our $3 + 1$ dimensional calculation in the ultrarelativistic limit. However, one should remark that this property is a feature of the particular choice of ultrarelativistic ideal gas equation of state \emph{and} our choice of the Landau hydrodynamic frame. Since (a) the necessary and sufficient causality conditions provided in proposition~\ref{prop:Lreal} and theorem~\ref{thm:LandauCausality} must hold for all values of $\psi\in[-1,1]$, and (b) the $D = 1 + 1$ case may be obtained by a particular choice of $\psi$, it follows that the $D = 1 + 1$ causality bounds represent a \emph{necessary} condition that must be satisfied for the causality in $D = 3 + 1$ to hold. For the ideal gas equation of state, Eq.~\eqref{eq:UR1Dcausality} is also sufficient for $D = 3 + 1$ nonlinear causality, meaning that the $1 + 1$ and $3 + 1$ constraints are equivalent in that case. However, we remark that the constraints coming from 1+1 analyses are generally not sufficient in higher spatial dimensions, which can be seen by the nontrivial angular dependence in the reality conditions in proposition~\ref{prop:Lreal}.

In addition to causality in the full nonlinear regime, a typical physical property one may want to ensure is that the current $J^\mu = nu^\mu + \mathcal{J}^\mu$ remains a future-pointing timelike vector throughout the evolution, in agreement with its interpretation as a baryon current. This is particularly relevant in astrophysical applications, where the number of baryons largely exceeds that of antibaryons and, thus, $n>0$. The condition for $J^\mu$ to be timelike (assuming of course $n\neq 0$) reads
\be
\label{eq:timelikeJB}
\left|\frac{\mathcal{J}}{n}\right| < 1 .
\ee
A quick look at the causality constraint in \eqref{eq:UR1Dcausality} shows that, depending on the transport coefficients,  $J^\mu$ can transition into a spacelike vector without violating causality. To illustrate this point, consider the choice of transport coefficient provided in \cite{HISCOCK1989125} derived from kinetic theory in the Landau frame \cite{ISRAEL1976310,ISRAEL1979341}
\be
\label{eq:LandauOmega}
\Omega_{\mathcal{J}}T = \frac{5\lambda}{4P}\left(\frac{\varepsilon + P}{n}\right)^2.
\ee
Here, $\lambda$ is a positive, constant parameter such that $\lambda = 1$ recovers the true ultrarelativistic limit.\footnote{The extra factor of $(\varepsilon + P)^2/n^2$ differs from \cite{HISCOCK1989125} since their work opts to write $\mathcal{J}^\mu = -n\bar{q}^\mu/(\varepsilon + P)$ at the level of the entropy current, meaning that their transport coefficients and relaxation equations are derived for $\bar{q}^\mu$ instead. We remark that letting $\mathcal{J}/n = \bar{q}/(\varepsilon + P)$ (such that $\bar{q}^2 = \bar{q}^\alpha\bar{q}_\alpha$), $\psi = +1$ and applying Eq.~\eqref{eq:LandauOmega} to Eq.~\eqref{eq:LandauDet} provides us the exact same determinant calculated in Eq.~(12) of \cite{HISCOCK1989125}, off by an extra factor of $x^6$ due to the extra spatial dimensions of our calculation.} Furthermore, for $\lambda = 1$, applying Eq.~\eqref{eq:LandauOmega} to our $1+1$ dimensional causality bound in Eq.~\eqref{eq:UR1Dcausality} recovers the constraint found in \cite{HISCOCK1989125}
\be
\label{eq:lambda1}
\frac{8}{5} \geq \left|\frac{\mathcal{J}}{n}\right|,
\ee
meaning that there exists a (nonlinearly) causally allowed region $\frac{8}{5} \geq \left|\frac{\mathcal{J}}{n}\right| \geq 1$ where $J^\mu$ is nontimelike. While Eq.~\eqref{eq:timelikeJB} ensures that the baryon current remains timelike, we emphasize that the condition $|\mathcal{J}/n| < 1$ is not a fundamental constraint from relativity or quantum field theory. In principle, configurations with $|\mathcal{J}/n| \geq 1$ could occur in systems with strong relative flows between different charge carriers. However, when $|\mathcal{J}/n|$ approaches or exceeds unity within Israel-Stewart theory, we are operating in a regime where the dissipative flux $\mathcal{J}^\mu$ is comparable to the equilibrium charge density $n$. This challenges the underlying assumption of the second-order expansion in the entropy current, where higher-order terms (beyond $\mathcal{J}^2$) could become significant. Thus, while our causality analysis shows that the equations remain mathematically causal even when $|\mathcal{J}/n| > 1$, such regimes likely lie outside the domain of validity of the truncated Israel-Stewart formulation. As such, one should be cautious when drawing conclusions about the physicality of the theory in the large dissipation limit $\mathcal{J}\sim n$ where higher order corrections can become relevant. Thus, instead of attributing physical meaning to causal evolution with spacelike baryon currents, we take this result to mean that nonlinear causality alone is not sufficient to guarantee that an effective theory is physically viable to a particular system of interest. Rather, one should use it in tandem with other fundamental properties, such as stability and local well-posedness \cite{ChoquetBruhatGRBook} to cite a few, to gain a more complete understanding about the true regime of validity of a theory. 

\subsection{Linear causality conditions}

A useful limiting case to consider when studying nonlinear causality is to instead investigate linear perturbations of the dynamic variables around some global equilibrium state. This procedure provides the \emph{linearized} equations of motion, from which we derive \emph{linear} conditions governing causality in an analogous manner to the nonlinear case. Conditions for causality using the general Israel-Stewart equations including both bulk and shear viscosity have been derived for the Eckart frame in \cite{Hiscock_Lindblom_stability_1983} and the Landau frame in \cite{Olson:1989ey}. However, we remark that the following linearized constraints were not derived explicitly in either work, but are equivalent to requiring that the characteristic wave speeds (roots of the characteristic determinant) are real and bounded by the speed of light, as argued in both papers (without specifying all of the roots with an explicit expression). Thus, our conditions are equivalent to those in the original works (without including bulk or shear viscosity). 

In a linearized analysis, one proceeds by writing the dynamic degrees of freedom in Eq.~\eqref{eq:Ldynamics} as
\be
\label{linear1}
\Umb = \Umb_{\textrm{eq.}} + \delta\Umb + \mathcal{O}(\delta\Umb^2),
\ee
where the ``eq." subscript refers to the equilibrium values and $\delta\Umb$ the linear fluctuations (for example, $\mathcal{J}^\mu_{\textrm{eq.}}=0$ but $\varepsilon_{\textrm{eq.}}$ and $n_{\textrm{eq.}}$ are nonzero). Applying Eq.~\eqref{linear1} to Eq.~\eqref{eq:LPDE}, we obtain the linearized equations of motion
\be
\mathbb{A}_{L,\textrm{eq.}}^\alpha\partial_\alpha\delta\Umb + \delta\left(\mathbb{B}_{L}\Umb\right)\bigg|_{\Umb = \Umb_{\textrm{eq.}}} +\mathcal{O}(\delta\Umb^2) = \boldsymbol{0}.
\ee
The corresponding characteristic determinant takes the form
\be
\det\left(\mathbb{A}_{L,\textrm{eq.}}^\alpha\phi_\alpha\right) = \Omega_{\mathcal{J}}^4\hat{x}^6v^{10}\left[\hat{x}^4 -\hat{x}^2 \left(c_s^2 +\frac{\Theta_n}{\Omega_\mathcal{J}}\right) + \frac{c_s^2 \Theta_n-\Theta_{\varepsilon|\bar{s}}P_n}{\Omega_{\mathcal{J}}}\right].
\ee
where we assume that all quantities above are evaluated at equilibrium besides the characteristic vector $\partial^\mu\delta\Umb\rightarrow \phi^\mu$, which instead corresponds to the linear fluctuations of the degrees of freedom $\Umb$ (denoted $\delta\Umb$) instead of $\Umb$ itself. Note that the characteristic determinant is now a quadratic in $\hat{x}^2$ for a fully arbitrary equation of state. One can immediately write down causality constraints that enforce that the roots (I) are bounded such that $0\leq \hat{x}^2\leq 1$ and (II) are real. Condition (I) can be found by applying theorem~\ref{thm:polysignchange} in  Appendix~\ref{App:Polynomials}, and (II) is just the condition for non-negativity of the discriminant. In summary, we find the following necessary and sufficient conditions
\bml
\label{eq:linearlandaucausality}
\bea
1\geq \frac{1}{2}\left(c_s^2 +\frac{\Theta_n}{\Omega_\mathcal{J}}\right) &\geq & 0,\\
\frac{c_s^2 \Theta_n-\Theta_{\varepsilon|\bar{s}}P_n}{\Omega_{\mathcal{J}}} &\geq & 0,\\
(1 - c_s^2)\left(1 - \frac{\Theta_n}{\Omega_{\mathcal{J}}}\right) - P_n\frac{\Theta_{\varepsilon|\bar{s}}}{\Omega_{\mathcal{J}}} &\geq & 0,\\
4P_n\frac{\Theta_{\varepsilon|\bar{s}}}{\Omega_{\mathcal{J}}} + \left(c_s^2 - \frac{\Theta_n}{\Omega_{\mathcal{J}}}\right)^2 &\geq & 0.
\eea
\eml
It is immediate to note that these conditions are remarkably simpler than the nonlinear constraints derived in proposition~\ref{prop:Lreal} and theorem~\ref{thm:LandauCausality}. More importantly, we note that these conditions fail to provide information on the dissipative contribution to the number current $\mathcal{J}^\mu$. We also remark that, in the ultrarelativistic limit considered in the previous section with $\Omega_{\mathcal{J}}$ defined via Eq.~\eqref{eq:LandauOmega}, these linear causality constraints simplify even further to a single, nonredundant bound on the transport coefficient:
\be
\lambda \geq \frac{1}{5}.
\ee
It should be clear then that, for \emph{any} fixed value of $\lambda$, linearized causality fails to provide a nontrivial causality bound on the dissipative current, which contrasts with the nonlinear case where $\mathcal{J}/n$ follows Eq.~\eqref{eq:lambda1}. Thus, we see that nonlinear analyses can provide useful, physical constraints on nonideal currents themselves, while linear analyses only constrain the transport coefficients and equilibrium values of the degrees of freedom at hand.


\section{Causality in the Eckart Frame}\label{sec:EckartFrame}

In the Eckart (or particle) frame the particle number current is parallel the fluid four-velocity. Here, dissipation is described by the energy diffusion vector $q^\mu$, such that $q^\mu u_\mu=0$. The conserved charges then take the form of Eq.~\eqref{eq:EckartConservedCharges}, which provides energy $u_\alpha\nabla_\beta T^{\alpha\beta} = 0$, momentum $\tensor{\Delta}{^\mu_\alpha}\nabla_\beta T^{\alpha\beta} = 0^\mu$, and baryon number $\nabla_\alpha J^\alpha = 0$ conservation equations in the form
\bml
\label{Eq_EM_Eckart}
\bea
\label{eq:Emomentum}
0 &=& \left[(\varepsilon +P)u^\alpha\tensor{g}{^\mu_\nu} + q^\alpha\tensor{g}{^\mu_\nu} + q^\mu\tensor{g}{^\alpha_\nu} - u^\alpha u^\mu q_\nu\right]\nabla_\alpha u^\nu + \Delta^{\alpha \mu}\nabla_\alpha P + u^\alpha\tensor{g}{^\mu_\nu}\nabla_\alpha q^\nu,\\
\label{eq:Eenergy}
0 &=& u^\alpha\nabla_\alpha \varepsilon + \left[(\varepsilon +P)\tensor{g}{^\alpha_\nu} + u^\alpha q_\nu\right]\nabla_\alpha u^\nu + \tensor{g}{^\alpha_\nu}\nabla_\alpha q^\nu,\\
\label{eq:Ebaryon}
0 &=& n\nabla_\alpha u^\alpha + u^\alpha\nabla_\alpha n.
\eea
\eml
In phenomenological Israel-Stewart theory, the nonideal dynamics are given by imposing a non-negative rate of entropy generation $T\nabla_\alpha\mathcal{S}^\alpha\geq 0$ on the entropy current
\begin{align}
\label{eq:Eentropycurrent}
T\mathcal{S}^\mu &= q^\mu + \left(Ts - \frac{\tau_q}{\kappa_q T}\frac{q^2}{2}\right)u^\mu + \mathcal{O}(q^3)
\end{align}
where $\tau_q$ is the relaxation time, $\kappa_q$ the thermal conductivity, and $q^2 \equiv q_\alpha q^\alpha$. 
The resulting Israel-Stewart relaxation equation describing energy diffusion is then
\be
\label{eq:Ediffusion}
0 = \frac{q^\mu}{\kappa_q T^2} + \frac{\Delta^{\mu\alpha}\nabla_\alpha T}{T^2} + \frac{1}{T}u^\alpha\nabla_\alpha u^\mu + \Omega_qu^\alpha\left(\nabla_\alpha q^\mu - u^\mu q_\nu\nabla_\alpha u^\nu\right) + \frac{1}{2}q^\mu \nabla_\alpha\left(\Omega_q u^\alpha\right),
\ee
where $\Omega_q \equiv {\tau_q}/{\kappa_q T^2}$.
Note that the last term in the above equation containing gradients of the transport coefficients is not present in the original literature \cite{ISRAEL1976310,ISRAEL1979341}. Here, we include it as it in general cannot be set to zero \textit{a priori} except in special cases \cite{RezzollaZanottiBookRelHydro}. Furthermore, above, we have applied the orthogonality relationship $u^\alpha q_\alpha = 0$ to ensure that the orthogonality constraint is propagated throughout the solution.

\subsection{The characteristic determinant}

As in the Landau frame, the equations of motion in Eq.~\eqref{eq:Eenergy}--\eqref{eq:Ebaryon},~\eqref{eq:Ediffusion} can be cast into a first-order quasilinear PDE of the form
\be
\label{eq:EPDE}
\mathbb{A}_E^\alpha\partial_\alpha\Umb + \mathbb{B}_E\Umb = \boldsymbol{0},
\ee
where we redefine the vector of degrees of freedom $\Umb\in\mathbb{R}^{10}$ in terms of the new dissipative flux $q^\mu$
\be
\label{eq:Edynamics}
\Umb \equiv
\begin{pmatrix}
u^\nu\\
q^\nu\\
\varepsilon\\
n
\end{pmatrix}.
\ee
As before, $\mathbb{A}_E^\mu$ and $\mathbb{B}_E$ are real-valued, $10\times 10$ matrices whose entries are nonlinear in terms of the components of $\Umb$ but not on their derivatives. Given the characteristic vector $\partial_\mu \Umb\rightarrow \phi_\mu$, the principal part of the theory takes the form
\be
\label{eq:Eprinciplepart}
\mathbb{A}_E^\alpha\phi_\alpha=
\begin{pmatrix}
[(\varepsilon + P)x + y_q]\tensor{g}{^\mu_\nu} + q^\mu\phi_\nu - xu^\mu q_\nu& x\tensor{g}{^\mu_\nu}& P_\varepsilon v^\mu& P_n v^\mu\\
\frac{x}{\Omega_qT}\tensor{g}{^\mu_\nu} + \frac{1}{2}q^\mu\phi_\nu - x u^\mu q_\nu& x\tensor{g}{^\mu_\nu}& \frac{T_\varepsilon v^\mu}{\Omega_qT^2} + \frac{\Omega_{q,\varepsilon} x}{2\Omega_q}q^\mu& \frac{T_n v^\mu}{\Omega_q T^2} + \frac{\Omega_{q,n} x}{2\Omega_q}q^\mu\\
(\varepsilon + P)\phi_\nu + xq_\nu& \phi_\nu& x& 0\\
n\phi_\nu& 0_\nu& 0& x
\end{pmatrix}.
\ee
Here, we write $y_q\equiv q^\alpha\phi_\alpha$, noting that it can be written in terms of $\psi\in[-1,1]$ via $y_q = q^\alpha v_\alpha = qv\psi$ since $q^\mu$ and $v^\mu$ are orthogonal to $u^\mu$ and, thus, $\Delta^{\mu\nu}$ defines an inner product over the space containing them. Proceeding forward with the calculation of the determinant immediately provides us with a more complex structure, including an order $5$ polynomial $P_5^{(E)}(\hat{x})$ and a nontrivial linear term
\begin{equation}
\label{eq:EckartDet}
\det(\mathbb{A}_E^\alpha\phi_\alpha) = E_q^3\left(E_q + \frac{\Omega_{q,\varepsilon}}{\Omega_q} q^2\right)v^{10}\hat{x}^3\left(\hat{x} + \frac{\psi q}{E_q}\right)^2P_5^{(E)}(\hat{x};\psi),\\
\end{equation}
where
\begin{equation}
\label{E_q_def}
    E_q \equiv \varepsilon+P-\frac{1}{\Omega_qT}
\end{equation}
and the order 5 polynomial $P_5^{(E)}$ can be expressed symbolically in the form
\begin{align}
P_5^{(E)}(\hat{x};\psi) &= \sum_{a = 0}^5\mathcal{A}_a^{(E)}(\psi)\hat{x}^a,
\end{align}
with the coefficients given by $\mathcal{A}_5^{(E)} = 1$ along with
\bml
\label{eq:ECoeffs}
\bea
\mathcal{A}_4^{(E)}(\psi) &=& \psi q \left[\frac{1}{2}\frac{3  +\frac{n\Omega_{q,n}}{\Omega_q}+ 2\left(\frac{T_\varepsilon}{\Omega_qT^2}- P_\varepsilon\right)}{E_q + \frac{\Omega_{q,\varepsilon}}{\Omega_q} q^2 }+\frac{1}{E_q}\right],\\
\mathcal{A}_3^{(E)}(\psi) &=& \frac{-\psi^2q^2}{E_q + \frac{\Omega_{q,\varepsilon}}{\Omega_q} q^2}\left[\frac{\Omega_{q,\varepsilon}}{\Omega_q} + \frac{\frac{\Omega_{q,\varepsilon}}{\Omega_q} \left(\frac{1}{\Omega_qT}\frac{n T_n}{T}-n P_n\right) - \left(3+\frac{n\Omega_{q,n}}{\Omega_q}\right) \left(\frac{1}{2} + \frac{1}{\Omega_qT}\frac{T_\varepsilon}{T}-P_\varepsilon\right)}{E_q}\right]\notag\\
&&\quad + \frac{\left(1 + \frac{n\Omega_{q,n}}{\Omega_q}\right)\left(P_\varepsilon - \frac{1}{\Omega_qT}\frac{T_\varepsilon}{T}\right) +\frac{\Omega_{q,\varepsilon}}{\Omega_q}\left(\frac{1}{\Omega_qT}\frac{nT_n}{T} - nP_n\right)}{\frac{\Omega_{q,\varepsilon}}{\Omega_q}E_q}\notag\\
&&\quad +\frac{P_\varepsilon\frac{\Omega_{q,\varepsilon}^2}{\Omega_q^2} q^2 - \left(1+\frac{n\Omega_{q,n}}{\Omega_q}\right)\left(P_\varepsilon-\frac{1}{\Omega_qT}\frac{T_\varepsilon}{T}\right)}{\frac{\Omega_{q,\varepsilon}}{\Omega_q}\left(E_q + \frac{\Omega_{q,\varepsilon}}{\Omega_q} q^2 \right)} - P_\varepsilon,\\
\mathcal{A}_2^{(E)}(\psi) &=& \frac{-\psi q}{E_q + \frac{\Omega_{q,\varepsilon}}{\Omega_q}q^2}\left[\frac{P_\varepsilon}{2} \left(1 + \frac{n\Omega_{q,n}}{\Omega_q}\right) + \frac{2}{\Omega_qT}\frac{T_\varepsilon}{T} - \frac{nP_n}{2}\frac{\Omega_{q,\varepsilon}}{\Omega_q} + \frac{n P_n-\frac{1}{\Omega_qT}\frac{nT_n}{T} + \psi^2 q^2\frac{\Omega_{q,\varepsilon}}{\Omega_q}}{E_q}\right],\\
\mathcal{A}_1^{(E)}(\psi) &=& \frac{1}{E_q + \frac{\Omega_{q,\varepsilon}}{\Omega_q} q^2}\left[\frac{1}{\Omega_qT}\left(nP_n \frac{T_\varepsilon}{T} - \frac{nT_n}{T}P_\varepsilon\right) +\frac{\psi^2q^2}{2}\frac{P_\varepsilon \left(1-\frac{n\Omega_{q,n}}{\Omega_q}\right)+\frac{\Omega_{q,\varepsilon}}{\Omega_q} n P_n -\frac{4}{\Omega_qT}\frac{T_\varepsilon}{T}}{E_q}\right],\\
\mathcal{A}_0^{(E)}(\psi) &=& \frac{\psi q}{\Omega_qT}\frac{nP_n \frac{T_\varepsilon}{T} - P_\varepsilon \frac{nT_n}{T}}{E_q \left(E_q + \frac{\Omega_{q,\varepsilon}}{\Omega_q} q^2 \right)}.
\eea
\eml
Direct comparison of \eqref{eq:ECoeffs} with the corresponding Landau frame result  in Eq.~\eqref{eq:LCoeffs} shows the significantly more complicated structure of the characteristic determinant and the explicit form of the coefficients found in the Eckart case. This result is not too surprising, as a choice of hydrodynamic frame imposes a change in the definition of the dynamics of the system, so one should not expect the general underlying structure to be preserved in the nonlinear case. Note that instead of a quartic such as $P_4^{(L)}$, the polynomial $P_5^{(E)}$ is quintic and, thus, Galois' famous result tell us that there is no general formula using only arithmetic operations and radicals to analytically solve this quintic equation in general. This difference in the structure of the characteristic determinant naturally has an important effect on reality conditions of the roots, which causality relies on.

\subsection{Nonlinear causality in the Eckart frame}\label{sec:NCinEckart}

Following the same strategy as in Landau's frame, we employ Eqs.~\eqref{Eq_EM_Eckart} and \eqref{eq:Ediffusion} as an \emph{extended system of equations}, where $u^0$ and $q^0$ are treated as dynamical variables. This extension serves as a mathematical tool to establish causality for solutions $(\varepsilon, n, u^\nu, q^\nu)$ of the original system, which are obtained by enforcing the constraints $u^\mu u_\mu + 1 = u_\nu q^\nu = 0$. The key observation is that any solution of the original system automatically satisfies the extended equations, but the converse is not true---the extended system admits additional, unphysical solutions. However, these extra solutions are irrelevant for our purposes, as we are solely interested in the causal propagation of the constrained solutions. The detailed justification for this approach can be found in the discussion following Eq.~\eqref{eq:causalitydef}.

In the case of Landau, the quartic nature of Eq.~\eqref{eq:LandauDet} allowed us to derive a set of simultaneously necessary \emph{and} sufficient conditions for causality, giving us an exact region where causality was satisfied. Although it is possible to derive separate sets of sufficient and necessary conditions for causality in the Eckart frame, which we provide here, the entire region is difficult to specify due to the inability to solve for the roots by analytic methods. In order to guarantee the number of real roots, one could impose conditions using Sturm's theorem \cite{SturmTheorem}, or equivalently note that the chain of conditions
\be
\forall\: 1\leq k\leq 4;\quad \left|\mathcal{A}_{k}^{(E)}\right|^2 - 4\left|\mathcal{A}_{k+1}^{(E)}\right|\left|\mathcal{A}_{k-1}^{(E)}\right| >0
\ee
provides a set of sufficient conditions guaranteeing that all roots of $P_5^{(E)}$ are (a) distinct and (b) exist in $\mathbb{R}$ \cite{SuffCondRealRoots}. Note that this condition is not necessary for reality of the roots. These constraints are not illuminating to write explicitly, so we suppress them here. However, provided the roots are real, one can provide necessary and sufficient conditions for causality. We present these below:
\begin{proposition}\label{prop:EckartCausality}
Suppose that all roots of $P_5^{(E)}$ are real (but possibly not distinct). The system $(\mathbb{A}_E^\alpha\partial_\alpha + \mathbb{B}_E)\normalfont{\Umb} = \boldsymbol{0}$ is causal if and only if the following conditions are satisfied for all $\psi\in[-1,1]$:
\bml
\bea
\left|\frac{q}{E_q}\right| &\leq & 1,\\
\frac{1}{5}\left|\mathcal{A}_4^{(E)}(\psi)\right| &\leq & \min\left\{1,2+\frac{1}{5}\mathcal{A}_3^{(E)}(\psi)\right\},\\
\frac{1}{10}\left|\mathcal{A}_2^{(E)}(\psi) + 6\mathcal{A}_4^{(E)}(\psi)\right| &\leq & 1 + \frac{3}{10}\mathcal{A}_3^{(E)}(\psi),\\
\frac{2}{5}\left|\mathcal{A}_2^{(E)}(\psi) + 2\mathcal{A}_4^{(E)}(\psi)\right| &\leq & 1 + \frac{1}{5}\mathcal{A}_1^{(E)}(\psi) + \frac{3}{5}\mathcal{A}_3^{(E)}(\psi),\\
\left|\mathcal{A}_0^{(E)}(\psi) + \mathcal{A}_2^{(E)} (\psi)+ \mathcal{A}_4^{(E)}(\psi)\right| &\leq & 1 + \mathcal{A}_1^{(E)}(\psi) + \mathcal{A}_3^{(E)}(\psi).
\eea
\eml
\begin{proof}
The proof follows the same steps as theorem~\ref{thm:LandauCausality}. First, one shows that the causality constraints enforce that all roots of $\det(\mathbb{A}_E^\alpha\phi_\alpha)$ must lie in $[-1,1]$, and then one shows that the above conditions are equivalent to imposing this constraint on the roots. The details of the proof can be found in Appendix~\ref{App:ECausality}.
\end{proof}
\end{proposition}

\subsection{Ultrarelativistic ideal gas}

As discussed in Sec.~\ref{subsec:URLandau}, for an ideal gas with an ultrarelativistic equation of state, our analysis recovers that of \cite{HISCOCK1989125}, in which the corresponding diffusion of matter (e.g. the baryon current) transitioned into a spacelike vector while still abiding by the corresponding nonlinear causality conditions. We saw that this strange behavior remains present also in a full $D = 3+ 1$ analysis in the Landau frame. An immediate question that arises is whether or not an analogous behavior occurs in the Eckart frame. 

In the Eckart frame, diffusion is taken into account in the energy-momentum tensor instead of the baryon current. Here, the baryon current remains timelike since $J^\mu\propto u^\mu$ by assumption. Thus, one must select a different yet analogous quantity to describe diffusion. Some useful historical connections are the \emph{energy conditions} (see, e.g. \cite{WaldBookGR1984}), which are physically motivated constraints imposed on (classical) relativistic theories related to the energy flow as viewed by nonspacelike observers. 

In this section, we work analogously to \cite{Hiscock_Lindblom_pathologies_1988}, which carries out the case of a $D = 1 + 1$ ultrarelativistic ideal gas in the Eckart frame. In this section, we show that (i) for a particular orientation of the energy diffusion vector $q^\mu$, our $D = 3 + 1$ calculation recovers the $D = 1 + 1$ calculation calculated in \cite{Hiscock_Lindblom_pathologies_1988}, and (ii) in contrast to the Landau frame case, we are unable to translate the relevant necessary and sufficient $D = 1 + 1$ causality conditions to the $D = 3 + 1$ case simply due to the change in structure of the characteristic determinant from a quartic (solvable in roots) to a quintic (not solvable by roots) as mentioned in Sec.~\ref{sec:NCinEckart}. This inability to completely specify nonlinear causality is therefore a salient difference between the Eckart and Landau frames that appears in a nonlinear analysis. 

Finally, in Sec.~\ref{subsec:DEC}, we draw an analogy to the Landau frame case in Sec.~\ref{subsec:URLandau} by (iii) showing that nonlinear causality is broken earlier than the \emph{dominant energy condition} in the $D = 1 + 1$ ultrarelativistic case (a necessary condition for $D = 3 + 1$), which contrasts conceptually with the Landau frame case in which nonlinear causality broke \emph{after} the transition of the baryon current to a spacelike vector.

We proceed in the same limit as referenced in \cite{Hiscock_Lindblom_pathologies_1988} by one again considering the ultrarelativistic ideal gas equation of state. As discussed in \cite{HISCOCK1989125}, we remark that \cite{Hiscock_Lindblom_pathologies_1988} used an incorrect form of the transport coefficient $\Omega_q$ which is analogous to Eq.~\eqref{eq:LandauOmega} except in units of energy diffusion (rather than number diffusion):
\be
\Omega_q\rightarrow \frac{5\lambda}{4P}.
\ee
This particular form of the transport coefficient is derived for the Landau frame \cite{ISRAEL1976310,ISRAEL1979341}, but can be transformed into the Eckart hydrodynamic frame up to linear order in disturbances of the fields from equilibrium $\mathcal{O}(\delta\Umb)$, via the transformation \cite{ISRAEL1976310,ISRAEL1979341}
\be
\Omega_q = \left(\frac{n}{\varepsilon + P}\right)^2\Omega_{\mathcal{J}} + \frac{1}{T(\varepsilon + P)} + \mathcal{O}(\delta\Umb).
\ee
We note that the corresponding relationships between the second order transport coefficients derived in Eq.~(15) of \cite{ISRAEL1976310} and Eq.~(2.42) of \cite{ISRAEL1979341} are dimensionally incorrect by a factor of $T$. Furthermore, we notice that the $\Omega$ transport coefficients used in this paper are related to $\beta$ and $\bar{\beta}$ in \cite{ISRAEL1976310,ISRAEL1979341} by $\Omega_q = \bar{\beta}/T$ and $\left(n/(\varepsilon + P)\right)^2\Omega_{\mathcal{J}} = \beta/T$. Fortunately, in the ultrarelativistic limit considered here, this transformation merely shifts the coefficient $\lambda\rightarrow\widetilde{\lambda}$ such that
\be
\label{eq:EckartURTransport}
\Omega_q T = \frac{5\widetilde{\lambda}}{4P};\quad \widetilde{\lambda} \equiv \lambda + \frac{1}{5}.
\ee
In this case, the structure of the determinant remains identical to Eq.~\eqref{eq:EckartDet}, except with the changes reflected in the coefficients due to the equation of state. This determinant is
\bml
\bea
\det(\mathbb{A}_E^\alpha\phi_\alpha) &=& \left(\frac{4}{3}\varepsilon\right)^4\alpha_q^3\left(\alpha_q -\frac{3}{2} \frac{q^2}{\varepsilon^2}\right)v^{10}\hat{x}^3\left(\hat{x} + \frac{\psi q}{E_q}\right)^2P_5^{(E)}(\hat{x};\psi),\\
\alpha_q(\widetilde{\lambda}) &\equiv & \frac{3}{4}\frac{E_q}{\varepsilon} = 1-\frac{1}{5\widetilde{\lambda}}
\eea
\eml
with marginally simplified coefficients ($\mathcal{A}_5^{(E)} =1$ as before)
\bml
\bea
\mathcal{A}_4^{(E)}&=& \frac{q}{\varepsilon}\psi\frac{\left(\frac{15}{4}-2\alpha_q\right)\alpha_q- \frac{9}{8}\frac{q^2}{\varepsilon^2}}{\alpha_q\left(\alpha_q - \frac{3}{2}\frac{q^2}{\varepsilon^2}\right)},\\
\mathcal{A}_3^{(E)} &=& \frac{\frac{3}{8}\left(1 + 5\psi^2\right)\frac{q^2}{\varepsilon^2} - \left(1 - \frac{2}{3}\alpha_q\right)\alpha_q }{\alpha_q \left(\alpha_q - \frac{3}{2}\frac{q^2}{\varepsilon^2}\right)},\\
\mathcal{A}_2^{(E)} &=& \frac{q}{\varepsilon}\psi\frac{\frac{9}{8}\frac{q^2}{\varepsilon^2}\psi^2 -\frac{3}{4}+ 2\alpha_q(\alpha_q-\frac{3}{4})}{\alpha_q\left(\alpha_q - \frac{3}{2}\frac{q^2}{\varepsilon^2}\right)},\\
\mathcal{A}_1^{(E)} &=& (1 - \alpha_q)\left(\frac{\frac{1}{3}\alpha_q-\frac{3}{2}\frac{q^2}{\varepsilon^2}\psi^2}{\alpha_q\left(\alpha_q - \frac{3}{2}\frac{q^2}{\varepsilon^2}\right)}\right),\\
\mathcal{A}_0^{(E)} &=& \psi \frac{q}{\varepsilon}\frac{1 - \alpha_q}{4\alpha_q\left(\alpha_q - \frac{3}{2}\frac{q^2}{\varepsilon^2}\right)}.
\eea
\eml
One can recover the exact determinant calculated in \cite{Hiscock_Lindblom_pathologies_1988} with $\lambda$ by setting $\psi = -1$ and then letting $\widetilde{\lambda}\rightarrow\lambda$, which then decouples the quintic polynomial into a quartic and a monomial. Therefore, our $D = 3 + 1$ calculation recovers the $D = 1 + 1$ case in the ultrarelativistic limit, as also shown in the Landau frame with \cite{HISCOCK1989125}. The corresponding determinant for $\psi = -1$ is
\begin{align}
\label{eq:HLdet}
\det(\mathbb{A}_E^\alpha\phi_\alpha)\bigg|_{\psi= -1} = &\left(\frac{4}{3}\varepsilon\right)^4\alpha_q^3\left(\alpha_q -\frac{3}{2} \frac{q^2}{\varepsilon^2}\right)v^{10}\hat{x}^3\left(\hat{x} - \frac{3}{4}\frac{q}{\varepsilon}\frac{1}{\alpha_q}\right)^3\notag\\
&\times\left[\hat{x}^4 - 2\frac{q}{\varepsilon}\frac{\frac{3}{2}-\alpha_q}{\alpha_q - \frac{3}{2}\frac{q^2}{\varepsilon^2}}\hat{x}^3+\frac{\frac{2}{3} \alpha_q + \frac{3}{2} \frac{q^2}{\varepsilon^2} - 1}{\alpha_q-\frac{3}{2}\frac{q^2}{\varepsilon^2}} \hat{x}^2 + \left(2\frac{q}{\varepsilon}\hat{x} + \frac{1}{3}\right)\left(\frac{1 - \alpha_q}{\alpha_q - \frac{3}{2} \frac{q^2}{\varepsilon^2}}\right)\right].
\end{align}
It is straightforward to show that substitution of $\alpha_q = 1 - 1/5\widetilde{\lambda}$ in terms of $\widetilde{\lambda}$ (rather than $\lambda$) provides the same determinant as Eq.~{(8)} in \cite{Hiscock_Lindblom_pathologies_1988}, off by an overall scalar expression in terms of $\widetilde{\lambda}$ and $q/\varepsilon$. In contrast to the ultrarelativistic Landau frame determinant in Eq.~\eqref{eq:URLandauDet}, it is important to mention that simply applying the equation of state for an ultrarelativistic gas in the Eckart hydrodynamic frame without fixing $\psi$ (akin to remaining in $3 + 1$ dimensions) fails to decouple the order 5 polynomial in the determinant, whereas in the Landau frame, the nontrivial components of the characteristic determinant were determined by a quadratic with relatively simple coefficients. As order 5 polynomials are not solvable by roots, we are therefore unable to show that the $D = 1 + 1$ case provides anything more than a necessary condition for causality in the $D = 3 + 1$ case (as our nonlinear bounds require that the inequalities hold for all $\psi\in[-1,1]$). 

It is also noteworthy that the presence of a linear prefactor in the determinant in Eq.~\eqref{eq:HLdet} also provides a nonlinear causality bound not present in the original paper \cite{Hiscock_Lindblom_pathologies_1988}, which arises from the higher dimensionality assumed on the equations of motion at the start of the calculation
\be
\label{eq:linbound}
\left|\frac{q}{\varepsilon}\right| \leq \frac{4}{3}\left(1 - \frac{1}{5\widetilde{\lambda}}\right).
\ee
We emphasize that Eq.~\eqref{eq:linbound} provides a bound not present in $D = 1 + 1$ dimensions, which differs from the Landau frame in the sense that the $D = 1 + 1$ determinant maintained all nontrivial information about causality in $D = 3 + 1$. 

Even though we will discuss the dominant energy condition in greater detail in the next section, here we anticipate that this (necessary) causality constraint enforces that the dominant energy condition is necessary for nonlinear causality by itself in $D = 3 + 1$ for $1/5 < \widetilde{\lambda} \leq 2/5$. The remaining nonlinear causality bounds in the spirit of those in proposition~\ref{prop:Lreal} and theorem~\ref{thm:LandauCausality} may be derived in the same fashion as the Landau frame, since both are order 4 polynomials in $\hat{x}$. For the particular angle $\psi = 0$ (as mentioned earlier, fixing particular angles amounts to a necessary but possibly not sufficient condition for the general $D = 3 + 1$ case), the characteristic determinant reduces to a quadratic in $\hat{x}^2$, from which it may be shown that our nonlinear causality bounds continue to enforce the dominant energy condition in the form
\be
\label{eq:psizerobound}
\left|\frac{q}{\varepsilon}\right| \leq \frac{2}{3} \sqrt{\frac{2}{12-45 \widetilde{\lambda}}-\frac{2}{5 \widetilde{\lambda}}+\frac{4}{3}} \leq \frac{2}{3}\quad\Leftrightarrow\quad \frac{2}{5}\leq \widetilde{\lambda}\leq \frac{2}{5}(2 + \sqrt{2})
\ee
Notice that this bound includes the true ultrarelativistic case in which $\lambda = 1$ ($\widetilde{\lambda} = 6/5$). 

As Eqs.~\eqref{eq:linbound}--\eqref{eq:psizerobound} guarantee the adherence to the dominant energy condition for smaller values of $\widetilde{\lambda}$, a natural question arises about whether or not the ultrarelativistic ideal gas equation of state permits the dominant energy condition to ever be broken for some $\lambda > 0$ ($\widetilde{\lambda} > 1/5$). Returning to the case of $\psi = -1$, we find that solving the nonlinear causality bounds numerically for arbitrary $\widetilde{\lambda}$ enforces the (necessary) generic bound
\be
\label{eq:nconstraint}
\left|\frac{q}{\varepsilon}\right| < \frac{\sqrt{2}}{3}\simeq 0.471,
\ee
for \emph{all} values of $\widetilde{\lambda}> 1/5$ \cite{CordeiroGit2025}. 

Finally, we note that formal constraints in terms of the degrees of freedom for nonlinear hyperbolicity \cite{ReulaStrongHyperbolic} are not provided. Thus, a formal analysis of hyperbolicity needs to be be approached for both cases, including the general $D = 3 + 1$ regime. 

\subsection{Dominant energy condition}\label{subsec:DEC}

The \emph{dominant energy condition (DEC)} \cite{WaldBookGR1984} is a physically motivated constraint typically valid in the classical regime that asserts for any future-directed timelike observer $\xi^\mu$ (DI) the energy density in the observers frame $\xi_\alpha\xi_\beta T^{\alpha\beta}\geq 0$ should never be negative, and (DII) the energy current as viewed by the observer $t_\xi^\mu \equiv -\xi_\alpha T^{\alpha\mu}$ must remain timelike and future-directed. Condition (DI) by itself is known as the \emph{weak energy condition (WEC)}, and therefore, the DEC implies the WEC by construction. One can show that this definition is equivalent to the condition that
\be
\label{eq:DECeigenvalue}
\varepsilon \geq |P_a|
\ee
for any such $a = 1,2,3$ where $P_i$ are the principal pressures (e.g. spatial eigenvalues) of $T^{\mu}_\nu$ \cite{WaldBookGR1984}.

While energy conditions are not fundamental, and violations are known to exist (see \cite{Kontou_2020} for a useful summary of such cases), they exist as useful guidelines for where one can expect most reasonable theories to exist. It is also important to mention that the DEC provides constraints that are often much simpler than those observed in a standard nonlinear causality analysis, yet also constrain the out-of-equilibrium currents directly, unlike a linear analysis. Thus, energy conditions contain information about nonideal fluxes that standard linearization procedures fail to describe, while still remaining relatively simple to calculate. We emphasize, however, that although the DEC appears like a suitable substitute for nonlinear causality as defined earlier, it is an artificial criterion imposed on what is \emph{expected} of a theory, and is neither necessary, nor sufficient for nonlinear causality in general.

In the Eckart frame case, notice that the energy flux in the fluid frame $\xi^\mu = u^\mu$ can be expressed as
\be
t_u^\mu \equiv -u_\alpha T^{\alpha\mu} = \varepsilon u^\mu + q^\mu.
\ee
The DEC then implies that $t_u^\mu$ must be future-directed timelike vector, which gives
\be
\label{eq:timelikefluid}
\left|\frac{q}{\varepsilon}\right| \leq 1.
\ee
In particular, one can show for the Eckart frame with diffusion that the DEC in Eq.~\eqref{eq:DECeigenvalue} is equivalent to the algebraic conditions
\bml
\label{eq:DEC}
\bea
\label{eq:DEC1}
|P| &\leq & \varepsilon,\\
\label{eq:DEC2}
\left|\frac{q}{\varepsilon + P}\right| &\leq & \frac{1}{2}.
\eea
\eml
In some cases, these conditions are more stringent than those of Eq.~\eqref{eq:timelikefluid}, as they must hold for all timelike observers $\xi^\mu$, not just $u^\mu$. For example, for an ultrarelativistic fluid satisfying $P \equiv P(\varepsilon) = \varepsilon/3$, the conditions in Eq.~\eqref{eq:DEC} reduce to
\be
\label{eq:URDEC}
\left|\frac{q}{\varepsilon}\right| \leq \frac{2}{3}.
\ee
Comparing with Eq.~\eqref{eq:nconstraint} suggests that the dominant energy condition \cite{WaldBookGR1984} is never broken for an ultrarelativistic ideal gas in $1 + 1$ dimensions (which then implies that this holds for $D = 3 + 1$ as well). We remark that this adherence to the dominant energy condition contrasts with the analogous Landau frame case for which the corresponding nonlinear causality constraints allowed spacelike propagation of the number current. However, it is important to reiterate that these constraints are dependent on the equation of state, and adherence to energy conditions should not be expected for different choices even within the Eckart frame. 

\subsection{Linear causality conditions}

As done for the Landau frame, here we derive linear causality bounds for the linearized equations of motion in the Eckart frame. Expressions for the characteristic speeds (roots of the characteristic determinant) were previously outlined in \cite{Hiscock_Lindblom_stability_1983} (with causality enforcing that the speeds must be between $-1$ and $+1$), but here, we provide explicit necessary and sufficient constraints that, to the best of our knowledge, were not previously given in the literature. After expanding the hydrodynamic fields as in Eq.~\eqref{linear1}, the characteristic determinant of the corresponding linearized system of Eqs. \eqref{eq:EPDE} decouples into
\be
\det\left(\mathbb{A}_E^\alpha\phi_\alpha\right) = (\varepsilon + P)^4v^{10}\hat{x}^6\left[\hat{x}^4 - \hat{x}^2 \left(c_s^2 - \frac{1}{\Omega_qT}\frac{nT_n}{T(\varepsilon +P)}\right) + \frac{n P_n T_{\varepsilon|\bar{s}}-c_s^2 n T_n}{T^2 \Omega_q(\varepsilon +P)}\right] .
\ee
Causality bounds are derived analogously to Eq.~\eqref{eq:linearlandaucausality} due to the linearized determinant having an identical structure
\bml
\bea
1 \geq \frac{1}{2} \left(c_s^2-\frac{1}{\Omega_qT}\frac{n T_n}{T(\varepsilon +P)}\right) &\geq & 0,\\
\frac{1}{\Omega_qT}\frac{1}{\varepsilon + P}\left(n P_n \frac{T_{\varepsilon|\bar{s}}}{T}-c_s^2\frac{nT_n}{T}\right) &\geq & 0,\\
(1-c_s^2)\left(1 + \frac{1}{\Omega_qT}\frac{nT_n}{T(\varepsilon +P)}\right)+\frac{1}{\Omega_qT}\frac{nP_n}{\varepsilon +P}\frac{T_{\varepsilon|\bar{s}}}{T} &\geq & 0,\\
\left(c_s^2 + \frac{1}{\Omega_qT}\frac{n T_n}{T(\varepsilon +P)}\right)^2 -\frac{4}{\Omega_q T}\frac{nP_n}{\varepsilon + P}\frac{T_{\varepsilon|\bar{s}}}{T}&\geq & 0.
\eea
\eml
A key difference between the nonlinear causality conditions in proposition~\ref{prop:EckartCausality} and our linear conditions is that our linear constraints are necessary \emph{and} sufficient due to the decoupling of terms in the characteristic determinant. Thus, the linearized constraints have the advantage of specifying the entire region of (linear) causality. However, we note that the only new nonideal term present in the bounds is the transport coefficient $\Omega_q$, and thus, as with the Landau frame, the linear conditions fail to place any meaningful restriction on the dissipative current $q^\mu$ itself. Furthermore, we note similarly that the ultrarelativistic limit of the linearized conditions with $\varepsilon/3 = P(\varepsilon) = nT(\varepsilon,n)$ and $\Omega_qT = 5\widetilde{\lambda}/4P$ from Eq.~\eqref{eq:EckartURTransport} once again reduces the causality constraints to a single condition
\be
\widetilde{\lambda} \geq \frac{1}{5}.
\ee
As with the Landau frame, this result contrasts with our nonlinear constraints provided in Eqs.~\eqref{eq:linbound}--\eqref{eq:nconstraint}, which directly constrain $q/\varepsilon$. This special case once again shows that while one may gain a simple and concise understanding on causality and the regime of validity of the theory through a linear analysis, this leaves out important physical information concerning the interplay of equilibrium and out-of-equilibrium quantities (the dissipative fluxes) only obtainable through a nonlinear approach.

\section{Conclusion}\label{sec:Conclusion}

In this paper, we have presented the first nonlinear causality analysis of Israel-Stewart theory with an energy/number dissipation current in $3 + 1$ dimensions, with no assumptions on the spacetime geometry or equation of state. This analysis goes beyond linear analyses \cite{Hiscock_Lindblom_stability_1983,Olson:1989ey,Brito:2020nou} and previous nonlinear causality analyses provided for an ultrarelativistic ideal gas in $1 + 1$ dimensions for both Eckart and Landau frames \cite{HISCOCK1989125,Hiscock_Lindblom_pathologies_1988}. In the Landau frame, these causality constraints uniquely specified the entire region of nonlinear causality via a set of simultaneously necessary and sufficient conditions, whereas in the Eckart frame, the order 5 polynomial present in the characteristic determinant prevented such conditions from being determined analytically. Instead, for Eckart we provided conditions that were \emph{sufficient} for the characteristic wave speeds (roots) to be real, and then provided conditions that were necessary and sufficient for causality \emph{assuming that these roots were real} (as opposed to guaranteeing reality and causality at the same time). This discrepancy directly expressed the difference in structure for the characteristic roots for different hydrodynamic frames. In either case, we showed that our general constraints in $3 + 1$ reduced to those found in $1+1$ in the case of an ultrarelativistic ideal gas equation of state after a suitable choice of angle. 

Causality has been used for many years to constrain formulations of relativistic fluid dynamics and their applications. However, as with any theory built upon an expansion scheme (in the case of the phenomenological Israel-Stewart theory we used, such an expansion is done at the level of the entropy current), one must be careful when estimating the regime of validity of such an approach. Naturally, one expects that the theory should break down if the expansion parameters become large enough, but the exact point of breakdown is not built into the definition of causality. In other words, even if one has a solution of the full nonlinear set of equations of motion that respects causality, one should not necessarily expect that the solutions are physically viable without further review\footnote{A trivial example is the following. The nonlinear PDE for some scalar field $\phi(t,x)$ given  by $(\partial_t^2 - \partial_x^2 +\alpha \partial_t)\phi + \phi^2=0$ is hyperbolic and causal for $\alpha \in \mathbb{R}$. However, only for $\alpha >0$ the solutions remain stable. Therefore, though causality may be considered a necessary condition, it is certainly not a sufficient requirement for ensuring physical behavior.}. Thus, causality is only one of many other checks and tests on a theory that may be applied in tandem to more precisely determine where one should trust a given theory of relativistic fluid dynamics.

As a brief illustration of this concept, using the ultrarelativistic ideal gas equation of state in  \cite{HISCOCK1989125,Hiscock_Lindblom_pathologies_1988}, there existed a region of solutions in the Landau hydrodynamic frame that were nonlinearly causal, but allowed the total baryon current $J^\mu$ to cross over to become a spacelike vector, if the dissipation current $\mathcal{J}^\mu$ is large enough. We extended this derivation to the $D = 3 + 1$ case and showed that it was implied directly by a previous $D = 1 + 1$ calculation in \cite{HISCOCK1989125}. An analogous transition to spacelike dissipation of diffusion in the Eckart frame, perhaps unsurprisingly, was not preserved between hydrodynamic frames under equivalent equations of state/classes of transport coefficients. We showed that the dominant energy condition was always satisfied for an ultrarelativistic ideal gas satisfying the nonlinear causality conditions in the Eckart frame, which implied that transition of the rest frame energy current $-u_\alpha T^{\alpha\mu}$ from timelike to spacelike was not reached until the theory was far outside of the regime of nonlinear causality. For a summary of these subcases and their implications, a visual depiction of the mathematical relationship between them may be found below in Fig.~\ref{fig:flowchart}. Note that necessary and sufficient conditions for causality are provided for all but two subcases corresponding to the two (nonlinear) $D = 3 + 1$ Eckart frame calculations, in which only sufficient conditions are provided.

\begin{figure}[!htb]
    \centering
    \includegraphics[width=0.8\linewidth]{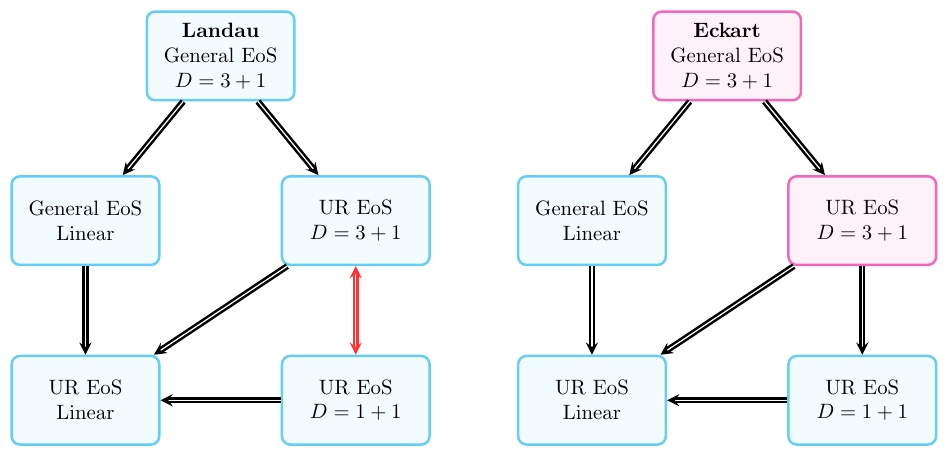}
    \caption{Flowchart depicting implications between subcases for diffusion in both Landau and Eckart frames. All boxes correspond to nonlinear constraints unless specified as linear. Boxes are color-coded by their adherence to the definition of causality prescribed by (CI) and (CII) in Eq.~\eqref{eq:causalitydef}---cyan signifies that the causality bounds provided in the paper are simultaneously necessary and sufficient, magenta signifies that only sufficient conditions are provided. Note that (A) $\Rightarrow$ (B) is read as ``(A) implies (B)," or equivalently, that ``(B) is necessary for (A)." Furthermore, ``(A) $\Leftrightarrow$ (B)" is read as ``(a) if and only if (B)" or ``(B) is necessary \emph{and} sufficient for (A)." For brevity, we write EoS = ``equation of state" and UR = ``ultrarelativistic (ideal gas)."}
    \label{fig:flowchart}
\end{figure}

Finally, as a brief check, we considered the regime of both frames linearized around equilibrium and verified that the linear causality conditions failed to directly constrain the out-of-equilibrium contributions from number/energy dissipation, as expected. In particular, the linear causality conditions for ultrarelativistic ideal gas cases in either frame reduced to a single constraint on the transport coefficient that failed to capture information about relevant energy conditions or the baryon current by construction.

Our work shows that one should be vigilant when testing the regime applicability of theories of relativistic hydrodynamics. Although causality may be considered to be a desired feature to be displayed by such effective theories, one should be cautious in drawing immediate conclusions about the suitability of a given theory even if causality holds. 

A possible extension of this work would be to analyze the nonlinear (strong) hyperbolicity \cite{ReulaStrongHyperbolic} properties of these theories of relativistic diffusion so that one can also discuss the existence and well-posedness properties of the solutions of their equations of motion. In the $D = 1 + 1$ case, sufficient numerical conditions for distinct roots are provided in \cite{HISCOCK1989125,Hiscock_Lindblom_pathologies_1988}, however, a formal nonlinear analysis of hyperbolicity, for both cases, particularly the more general $D = 3 + 1$ case is still necessary to ensure the local well-posedness of the Cauchy problem \cite{ChoquetBruhatGRBook} for both the Eckart and Landau cases. The provided nonlinear causality bounds may act as a useful guideline for this calculation to obtain sufficient conditions that ensures hyperbolicity. We leave this challenging problem to future work.

\section*{Acknowledgements}

We thank M.~Disconzi for discussions. E.S. has received funding from the European Union’s Horizon Europe research and innovation program under the Marie Sk\l odowska-Curie grant agreement No. 101109747. J.N. and I.C. are partly
supported by the U.S. Department of Energy, Office of Science, Office for Nuclear Physics under Award
No. DE-SC0023861. This work was done while the author F.S.B. was a Research Assistant Professor at Vanderbilt University.

\section*{DATA AVAILABILITY}

The data that support the findings of this article are openly available \cite{CordeiroGit2025}.

\def\cprime{$'$}
%


\appendix
\section{ELEMENTARY CONSTRAINTS ON POLYNOMIAL ROOTS}\label{App:Polynomials}

Here we derive necessary and sufficient constraints for the roots of polynomials to be constrained between $-1$ and $+1$, assuming that they exist in $\mathbb{R}$. theorem~\ref{thm:polysignchange} provides necessary and sufficient constraints for the roots of any polynomial of order $n\geq 1$ to be non-negative, and corollary~\ref{cor:quarticunity} provides constraints for the order $n = 4$ and $n = 5$ case such that the roots are instead bounded between $-1$ and $+1$, which follows directly from Theorem~\ref{thm:polysignchange}. Here are the results.

\begin{theorem}\label{thm:polysignchange}
Let $P_n$ be a real-valued polynomial with $n$ real roots $r_k$ where
\be
P_n(X) = \sum_{k = 0}^{n}\mathcal{A}_kX^k,
\ee
where $n\geq 1$, $n\in\mathbb{N}$ and $\mathcal{A}_k\in\mathbb{R}$ $(\mathcal{A}_n = 1)$. Then $\forall k\in\{1,2,\cdots,n\}$, $r_k\geq 0$ if, and only if the ordered $n$-tuple $(1,\mathcal{A}_{n-1},\mathcal{A}_{n-2},\dotso,\mathcal{A}_0)\in\mathbb{R}^{n}$ forms an alternating sequence with $\mathcal{A}_{n - 1}\leq 0$.
\begin{proof}
This proof is actually an elementary result related to Descartes' sign rule. We can prove it by using induction.\\

\noindent\textit{Base Case:} Let $n = 1$. Then, we consider the monomial $P_1(X) = X + \mathcal{A}_0$. By the fundamental theorem of algebra, we can also express it in terms of the roots as $X - r_1$. Equating the two provides $\mathcal{A}_0 = -r_1$. If $r_1\geq 0$, clearly, $\mathcal{A}_0\leq 0$, and vice versa, proving the ``if and only if" statement.\\

\noindent\textit{Induction Step:} Suppose that there exists $m > 1$ such that $P_m$ has $m$ real roots. Suppose that these roots are non-negative if, and only if $(1,\mathcal{A}_{m-1},\mathcal{A}_{m-2},\dotso,\mathcal{A}_0)\in\mathbb{R}^m$ forms an alternating sequence with $\mathcal{A}_{m - 1}\leq 0$. Consider instead the order $m + 1$ polynomial with $m + 1$ real roots. Notice that, by the fundamental theorem of algebra, we can express it as a monomial times an order $m$ polynomial
\begin{align}
P_{m+1}(X) &= (X - r_{m+1})P_m(X),\notag\\
&= (X - r_{m + 1})\sum_{k = 0}^m \mathcal{A}_{k}X^k,\notag\\
&= X^{m + 1} + \sum_{k = 1}^m (\mathcal{A}_{k - 1} - r_{m+1}\mathcal{A}_{k})X^{k} - r_{m +1}\mathcal{A}_0
\end{align}
where $\mathcal{A}_m = 1$. Therefore, we are considering the sequence
\be
(1,\widetilde{\mathcal{A}}_m,\widetilde{\mathcal{A}}_{m-1},\dotso,\widetilde{\mathcal{A}}_0) = (1,-r_{m+1}\mathcal{A}_m + \mathcal{A}_{m-1},-r_{m+1}\mathcal{A}_{m-1}+\mathcal{A}_{m-2},\dotso,-r_{m+1}\mathcal{A}_0).
\ee
Suppose that the roots are nonnegative. Notice that for any $k\in\{1,2,\dotso,m\}$, $\textrm{sgn}(\mathcal{A}_k) = -\textrm{sgn}(\mathcal{A}_{k-1})$. Therefore, $\textrm{sgn}(\widetilde{\mathcal{A}}_k = -r_{m+1}\mathcal{A}_k + \mathcal{A}_{k-1}) = \textrm{sgn}(\mathcal{A}_{k-1})$, since $r_{m+1}\geq 0$. In other words, $(1,\widetilde{\mathcal{A}}_m,\widetilde{\mathcal{A}}_{m-1},\dotso,\widetilde{\mathcal{A}}_0)$ is then an alternating sequence where $\widetilde{\mathcal{A}}_m\leq 0$.\\

Now, suppose conversely that $(1,\widetilde{\mathcal{A}}_m,\widetilde{\mathcal{A}}_{m-1},\dotso,\widetilde{\mathcal{A}}_0)$ is alternating with $\widetilde{\mathcal{A}}_m\leq 0$. Let us assume by contradiction that there exists some $j\in\{1,2,\dotso,m+1\}$ such that $r_j < 0$. Without loss of generality, we can assume this to be $j = m + 1$, since choosing a particular label for a root merely shifts around labels, which (in this case) are symmetric under interchange. It is easier to see from the unique factorization given by the fundamental theorem of algebra
\be
P_{m + 1}(X) = \prod_{k = 1}^{m+1}(X - r_k).
\ee
Notice that since $r_{m + 1} < 0$ is a root of the equation, we have
\begin{align}
0 &= \sum_{k = 0}^{m + 1}\widetilde{\mathcal{A}}_kr_{m + 1}^k,\notag\\
&= \sum_{k = 2j\textrm{ even}}\widetilde{\mathcal{A}}_{2j}r_{m + 1}^{2j} + \sum_{k = 2j + 1 \textrm{ odd}}\widetilde{\mathcal{A}}_{2j + 1}r_{m + 1}^{2j + 1},\notag\\
&= \sum_{k = 2j\textrm{ even}}\widetilde{\mathcal{A}}_{2j}|r_{m + 1}|^{2j} -\sum_{k = 2j + 1 \textrm{ odd}}\widetilde{\mathcal{A}}_{2j + 1}|r_{m + 1}|^{2j + 1}.
\end{align}
We shall consider two cases. Suppose that $m + 1$ is even. Then all odd coefficients $|\widetilde{\mathcal{A}}_{2j + 1}| = -\widetilde{\mathcal{A}}_{2j + 1}$ are non-positive, and all even ones are non-negative. We obtain
\begin{align}
0 &= \sum_{j}|\widetilde{\mathcal{A}}_{2j}||r_{m + 1}|^{2j}  + \sum_{j}|\widetilde{\mathcal{A}}_{2j + 1}||r_{m + 1}|^{2j + 1},\\
&= \sum_{k = 0}^{m + 1}|\widetilde{\mathcal{A}}_{k}||r_{m + 1}|^{k}.
\end{align}
A finite sum of non-negative numbers that equals zero necessitates that each term in the sum is zero. However, since $r_{m + 1}\neq 0$, this means that for all $k$, $\widetilde{\mathcal{A}}_k = 0$, even though $\widetilde{\mathcal{A}}_{m + 1} = 1$. Hence, we have reached a contradiction. Now, instead, assume that $m + 1$ is odd. Then all even coefficients are nonpositive and all odd coefficients are non-negative:
\begin{align}
0 &= -\sum_{j}|\widetilde{\mathcal{A}}_{2j}||r_{m + 1}|^{2j}  - \sum_{j}|\widetilde{\mathcal{A}}_{2j + 1}||r_{m + 1}|^{2j + 1},\\
&= -\sum_{k = 0}^{m + 1}|\widetilde{\mathcal{A}}_{k}||r_{m + 1}|^{k}.
\end{align}
Multiplying each side by $-1$ yields an identical argument to the previous case. Therefore, in either case, we reach a contradiction, which shows that each root must be non-negative.\\

In conclusion, we have shown that the base $n = 1$ case is satisfied, and that for some $m > 1$, given that requirement that the coefficients of $P_m$ must alternate in sign is a necessary and sufficient condition for the non-negativity of the roots, it follows that the same holds for $P_{m + 1}$, assuming the reality of the roots. This result proves the statement by induction.
\end{proof}
\end{theorem}
Next, we prove a corollary of theorem~\ref{thm:polysignchange} that follows directly from the $n = 4,5$ cases. Instead of ensuring that the roots are non-negative, corollary~\ref{cor:quarticunity} instead provides conditions for the roots to be bounded between $-1$ and $+1$. 
\begin{corollary}\label{cor:quarticunity}
Let $P_n$ for $n\geq 1$ be a polynomial with all real roots $r_k$ and coefficients $\mathcal{A}_k$ for $k = 1,\dotso,n$. Consider the $n = 4$ case
\be
P_4(X) = X^4 + \mathcal{A}_3X^3 + \mathcal{A}_2X^2 + \mathcal{A}_1X + \mathcal{A}_0.
\ee
These two sets of conditions imply that the following are necessary and sufficient for the roots to satisfy $|r_k|\leq 1$
\bml
\bea
\frac{1}{4}\left|\mathcal{A}_3\right| &\leq & \min\left\{1, \frac{1}{2} + \frac{1}{12}\mathcal{A}_2\right\},\\
\frac{1}{4}\left|\mathcal{A}_1 + 3\mathcal{A}_3\right| &\leq & 1 + \frac{1}{2}\mathcal{A}_2,\\
\left|\mathcal{A}_1 + \mathcal{A}_3\right| &\leq & 1+\mathcal{A}_0 + \mathcal{A}_2.
\eea
\eml
Similarly, consider the quintic polynomial
\be
P_5(X) = X^5 + \mathcal{A}_4X^4 + \mathcal{A}_3X^3 + \mathcal{A}_2X^2 + \mathcal{A}_1X + \mathcal{A}_0.
\ee
The following conditions are necessary and sufficient for the polynomial roots to be bounded via $|r_k|\leq 1$ for all such $k$
\bml
\bea
\frac{|\mathcal{A}_4|}{5} &\leq & 1 ,\\
4|\mathcal{A}_4| &\leq & 10 + \mathcal{A}_3 ,\\
|\mathcal{A}_2 + 6\mathcal{A}_4| &\leq & 10 + 3\mathcal{A}_3 ,\\
|2\mathcal{A}_2 + 4\mathcal{A}_4| &\leq & 5 + \mathcal{A}_1 + 3\mathcal{A}_3,\\
|\mathcal{A}_0 + \mathcal{A}_2 + \mathcal{A}_4|&\leq & 1+ \mathcal{A}_1 + \mathcal{A}_3.
\eea
\eml
\begin{proof}
Let $z_\pm \equiv 1 \pm X$. Then if $z_+\geq 0$, then $X \geq -1$ and if $z_-\geq 0$, then $X \leq 1$. One can rewrite the polynomial $P_4$ as
\begin{align}
P_4(X = \pm (z_\pm - 1)) 
= z_\pm^4 &+ (-4\pm \mathcal{A}_3)z_\pm^3+ \left(6 \mp 3\mathcal{A}_3 + \mathcal{A}_2\right)z_\pm^2 + \left(-4\pm 3\mathcal{A}_3 - 2\mathcal{A}_2\pm \mathcal{A}_1\right)z_\pm\notag\\
& + \left(1+\mathcal{A}_0 \mp \mathcal{A}_3 + \mathcal{A}_2\mp \mathcal{A}_1\right).
\end{align}
Applying theorem~\ref{thm:polysignchange} (e.g. the coefficients of the polynomial must alternate in sign) to both the polynomial in $z_+$ and $z_-$, we get the bounds that are necessary and sufficient for $r_k \geq -1$
\bml
\bea
0 &\geq & -4 - \mathcal{A}_3,\\
0 &\geq & -4 - 3\mathcal{A}_3 - 2\mathcal{A}_2 - \mathcal{A}_1,\\
0 &\leq & 6 + 3\mathcal{A}_3 + \mathcal{A}_2,\\
0 &\leq & 1+\mathcal{A}_0 + \mathcal{A}_3 + \mathcal{A}_2 + \mathcal{A}_1,
\eea
\eml
and the set of necessary and sufficient conditions for $r_k \leq +1$ (for $z_-$)
\bml
\bea
0 &\geq & \mathcal{A}_3 - 4,\\
0 &\geq & -4 + 3\mathcal{A}_3 - 2\mathcal{A}_2 + \mathcal{A}_1,\\
0 &\leq & 6- 3\mathcal{A}_3 + \mathcal{A}_2,\\
0 &\leq & 1+\mathcal{A}_0 - \mathcal{A}_3 + \mathcal{A}_2 - \mathcal{A}_1.
\eea
\eml
The combined bounds on the roots $|r_k|\leq 1$ in the problem statement are found upon simple algebraic manipulation between the bounds of $r_k \leq +1$ together with those for $r_k \geq -1$. Next, we consider the quintic $P_5$ as defined in the statement of the corollary. One finds that
\begin{align}
P_5(X = \pm(z_\pm - a)) = z_\pm^5 &+ (-5 \pm \mathcal{A}_4)z_\pm^4 + (10 + \mathcal{A}_3 \mp 4\mathcal{A}_4)z_\pm^3 + (-10 \pm\mathcal{A}_2 - 3\mathcal{A}_3 \pm 6\mathcal{A}_4)z_\pm^2\notag\\
&+ (5 + \mathcal{A}_1\mp 2\mathcal{A}_2 + 3\mathcal{A}_3 \mp 4\mathcal{A}_4)z_\pm + (-1\pm\mathcal{A}_0 - \mathcal{A}_1 \pm\mathcal{A}_2 - \mathcal{A}_3 \pm\mathcal{A}_4).
\end{align}
Conditions for $r_k\geq -1$ are then
\bml
\bea
0 &\geq & -5 + \mathcal{A}_4,\\
0 &\leq & 10 + \mathcal{A}_3 - 4\mathcal{A}_4,\\
0 &\geq & -10 +\mathcal{A}_2 - 3\mathcal{A}_3 +6\mathcal{A}_4,\\
0 &\leq & 5 + \mathcal{A}_1- 2\mathcal{A}_2 + 3\mathcal{A}_3 - 4\mathcal{A}_4,\\
0 &\geq & -1+\mathcal{A}_0 - \mathcal{A}_1 +\mathcal{A}_2 - \mathcal{A}_3 +\mathcal{A}_4
\eea
\eml
and similarly for $r_k \leq 1$
\bml
\bea
0 &\geq & -5 - \mathcal{A}_4,\\
0 &\leq & 10 + \mathcal{A}_3 + 4\mathcal{A}_4,\\
0 &\geq & -10 -\mathcal{A}_2 - 3\mathcal{A}_3 - 6\mathcal{A}_4,\\
0 &\leq & 5 + \mathcal{A}_1+ 2\mathcal{A}_2 + 3\mathcal{A}_3 + 4\mathcal{A}_4,\\
0 &\geq & -1-\mathcal{A}_0 - \mathcal{A}_1 -\mathcal{A}_2 - \mathcal{A}_3 -\mathcal{A}_4.
\eea
\eml
Combining the two sets provides the result in the problem statement. This completes the proof.
\end{proof}
\end{corollary}


\section{PROOF OF NONLINEAR CAUSALITY IN THE LANDAU FRAME}\label{App:LCausality}

We wish to find the region where the system
\be
\left(\mathbb{A}_L^\alpha\phi_\alpha + \mathbb{B}_L\right)\Umb = 0
\ee
has solutions that propagate causally. The characteristic determinant has the structure
\be
\det(\mathbb{A}_L^\alpha\phi_\alpha) = v^{10}\hat{x}^6P_4^{(L)}(\hat{x};\psi)
\ee
where $\hat{x}\equiv x/v$ and the quartic polynomial is written symbolically as
\be
P_4^{(L)}(\hat{x};\psi) = \sum_{a = 0}^4\mathcal{A}_a^{(L)}(\psi) \hat{x}^a
\ee
where $\mathcal{A}_4^{(L)} = 1$ and $\psi\in[-1,1]$. We refer the reader to Eq.~\eqref{eq:causalitydef} for the formal statement of causality.

Suppose that proposition~\ref{prop:Lreal} holds. Then the roots of $\det(\mathbb{A}^\alpha_L\phi_\alpha) = 0$ exist in $\mathbb{R}$ and (CI) in Eq.~\eqref{eq:causalitydef} is true. The roots $x = 0$ are immediately causal by definition of causality, since they correspond to the condition $\phi_0 = 0$ in the local rest frame, meaning that $\phi_\alpha \phi^\alpha = \sum_{j = 1}^3\phi_i^2\geq 0$. Hence, it suffices to consider the nontrivial quartic instead. Since the roots are real, there exists $r_a = r_a(\psi)$ for $a = 1,2,3,4$ such that
\be
P_4^{(L)}(\hat{x};\psi) = \prod_{a = 1}^4(\hat{x} - r_a(\psi)).
\ee
Since the roots are given by $x = r_a$, we can shift to the local rest frame to get $\phi_0^2 = r_a^2\sum_{j = 1}^2\phi_j^4$. Since $\phi_\alpha$ is nontimelike $\phi_\alpha \phi^\alpha = -\phi_0^2 + \sum_{j = 1}^3\phi_i^2\geq 0$, and imposing this condition on the roots implies that $r_a^2\leq 1$, which is equivalent to the condition $|r_a|\leq 1$. Corollary~\ref{cor:quarticunity} provides a necessary and sufficient condition for all $a = 1,2,3,4$ to be bounded such that $|r_a|\leq 1$ in terms of coefficients of $P_4^{(L)}$. They take the form
\bml
\bea
\frac{1}{4}\left|\mathcal{A}_3^{(L)}\right| &\leq & \inf\left\{1, \frac{1}{2} + \frac{1}{6}\mathcal{A}_2^{(L)}\right\},\\
\frac{1}{4}\left|3\mathcal{A}_3^{(L)} + \mathcal{A}_1^{(L)}\right| &\leq & 1 + \frac{1}{2}\mathcal{A}_2^{(L)},\\
\left|\mathcal{A}_3^{(L)} + \mathcal{A}_1^{(L)}\right| &\leq & 1+  \mathcal{A}_0^{(L)} + \mathcal{A}_2^{(L)}.
\eea
\eml
We note that these constraints must hold for all values of $\psi\in[-1,1]$. Plugging in the coefficients of $P_4^{(L)}$ gives us the constraints
\bml
\bea
\frac{|\psi|\mathcal{J}}{4} \left|\frac{P_n}{\varepsilon + P} + \frac{\Omega_{\mathcal{J},n}}{2\Omega_{\mathcal{J}}}\right| &\leq & \inf\left\{1,\frac{1}{2}\left(1-\frac{c_s^2}{6}\right) - \frac{\Theta_n}{12\Omega_{\mathcal{J}}}\right\},\\
\frac{|\psi|\mathcal{J}}{4} \left|\frac{\Omega_{\mathcal{J},n}}{\Omega_{\mathcal{J}}}\left(3-c_s^2\right) + \frac{5P_n}{\varepsilon + P} + \frac{\Omega_{\mathcal{J},\varepsilon|\bar{s}}}{\Omega_{\mathcal{J}}}P_n\right| &\leq & \left(1-c_s^2\right) + \left(1-\frac{\Theta_n}{\Omega_{\mathcal{J}}}\right),\\
\frac{|\psi|\mathcal{J}}{2} \left|\frac{\Omega_{\mathcal{J},n}}{\Omega_{\mathcal{J}}}\left(1-c_s^2\right) + \frac{P_n}{\varepsilon + P} + \frac{\Omega_{\mathcal{J},\varepsilon|\bar{s}}}{\Omega_{\mathcal{J}}}P_n\right| &\leq & (1-c_s^2)\left(1 -\frac{\Theta_n}{\Omega_{\mathcal{J}}}\right) -\frac{P_n}{\Omega_{\mathcal{J}}}\Theta_{\varepsilon|\bar{s}}.
\eea
\eml
It also follows that the case where $\psi = 1$ is the most stringent, and therefore enables the inequalities above to be satisfied for all such $\psi$, meaning that one can equivalently write
\bml
\bea
\frac{\mathcal{J}}{4} \left|\frac{P_n}{\varepsilon + P} + \frac{\Omega_{\mathcal{J},n}}{2\Omega_{\mathcal{J}}}\right| &\leq & \inf\left\{1,\frac{1}{2}\left(1-\frac{c_s^2}{6}\right) - \frac{\Theta_n}{12\Omega_{\mathcal{J}}}\right\},\\
\frac{\mathcal{J}}{4} \left|\frac{\Omega_{\mathcal{J},n}}{\Omega_{\mathcal{J}}}\left(3-c_s^2\right) + \frac{5P_n}{\varepsilon + P} + \frac{\Omega_{\mathcal{J},\varepsilon|\bar{s}}}{\Omega_{\mathcal{J}}}P_n\right| &\leq & \left(1-c_s^2\right) + \left(1-\frac{\Theta_n}{\Omega_{\mathcal{J}}}\right),\\
\frac{\mathcal{J}}{2} \left|\frac{\Omega_{\mathcal{J},n}}{\Omega_{\mathcal{J}}}\left(1-c_s^2\right) + \frac{P_n}{\varepsilon + P} + \frac{\Omega_{\mathcal{J},\varepsilon|\bar{s}}}{\Omega_{\mathcal{J}}}P_n\right| &\leq & (1-c_s^2)\left(1 -\frac{\Theta_n}{\Omega_{\mathcal{J}}}\right) -\frac{P_n}{\Omega_{\mathcal{J}}}\Theta_{\varepsilon|\bar{s}}.
\eea
\eml
This completes the proof.


\section{PROOF OF NONLINEAR CAUSALITY IN THE ECKART FRAME}\label{App:ECausality}

Analogously to the Landau case, we wish to find the region where the system
\be
\left(\mathbb{A}_E^\alpha\phi_\alpha + \mathbb{B}_E\right)\Umb = 0
\ee
has solutions that propagate causally. Here, the characteristic determinant takes the form
\bml
\bea
\det(\mathbb{A}_E^\alpha\phi_\alpha) &=& E_q^3\left(E_q + \frac{\Omega_{q,\varepsilon}}{\Omega_q} q^2\right)v^{10}\hat{x}^3\left(\hat{x} + \frac{\psi q}{E_q}\right)^2P_5^{(E)}(\hat{x};\psi),\\
E_q &\equiv& \varepsilon+P-\frac{1}{\Omega_qT}
\eea
\eml
where $P_5^{(E)}$ can be expressed symbolically as
\begin{align}
P_5^{(E)}(\hat{x};\psi) &= \sum_{a = 0}^5\mathcal{A}_a^{(E)}(\psi)\hat{x}^a,
\end{align}
with $\hat{x}\equiv x/v$ and $\psi\in[-1,1]$. 

Suppose that these roots are all real (possibly nondistinct). Then (CI) in Eq.~\eqref{eq:causalitydef} holds by assumption. As mentioned in the main paper, there are a couple of ways to guarantee reality of the roots, but finding the entire region where all roots are real is not possible by method of roots. As before, the roots $x = 0$ are immediately causal. Rotating into the local rest frame such that $x\rightarrow\phi_0$ and $v^2\rightarrow \sum_{j= 1}^3\phi_j^2$ allows us to immediately constrain the linear term provided by $\hat{x} = -\psi q/E_q$ (which exists in $\mathbb{R}$ as long as $E_q\neq 0$) as:
\be
\left|\frac{\psi q}{E_q}\right| \leq 1.
\ee
Here, we used the spacelike condition on the characteristics (CII) in Eq.~\eqref{eq:causalitydef} in order to eliminate the characteristics as before in the Landau case. Note that $|\psi| = 1$ provides the strongest bound, so we can just write $|q/E_q|\leq 1$. Finally for the order $5$ polynomial $P_5^{(E)}$, one notes by the fundamental theorem of algebra that given the roots $r_k$ for $k = 1,\dotso,5$:
\be
P_5^{(E)}(\hat{x};\psi) = \prod_{k = 1}^5(\hat{x} - r_k(\psi))
\ee
and one can again impose the spacelike condition to get the bounds $|r_k|\leq 1$. Corollary~\ref{cor:quarticunity} once again tells us that these conditions on the roots are logically equivalent to constraints on the coefficients $\mathcal{A}_k^{(E)}$. That is
\bml
\bea
\frac{1}{5}\left|\mathcal{A}_4^{(E)}(\psi)\right| &\leq & \min\left\{1,2+\frac{1}{5}\mathcal{A}_3^{(E)}(\psi)\right\},\\
\frac{1}{10}\left|\mathcal{A}_2^{(E)}(\psi) + 6\mathcal{A}_4^{(E)}(\psi)\right| &\leq & 1 + \frac{3}{10}\mathcal{A}_3^{(E)}(\psi),\\
\frac{2}{5}\left|\mathcal{A}_2^{(E)}(\psi) + 2\mathcal{A}_4^{(E)}(\psi)\right| &\leq & 1 + \frac{1}{5}\mathcal{A}_1^{(E)}(\psi) + \frac{3}{5}\mathcal{A}_3^{(E)}(\psi),\\
\left|\mathcal{A}_0^{(E)}(\psi) + \mathcal{A}_2^{(E)} (\psi)+ \mathcal{A}_4^{(E)}(\psi)\right| &\leq & 1 + \mathcal{A}_1^{(E)}(\psi) + \mathcal{A}_3^{(E)}(\psi).
\eea
\eml
This completes the proof.


\begin{thebibliography}{65}%
	\makeatletter
	\providecommand \@ifxundefined [1]{%
		\@ifx{#1\undefined}
	}%
	\providecommand \@ifnum [1]{%
		\ifnum #1\expandafter \@firstoftwo
		\else \expandafter \@secondoftwo
		\fi
	}%
	\providecommand \@ifx [1]{%
		\ifx #1\expandafter \@firstoftwo
		\else \expandafter \@secondoftwo
		\fi
	}%
	\providecommand \natexlab [1]{#1}%
	\providecommand \enquote  [1]{``#1''}%
	\providecommand \bibnamefont  [1]{#1}%
	\providecommand \bibfnamefont [1]{#1}%
	\providecommand \citenamefont [1]{#1}%
	\providecommand \href@noop [0]{\@secondoftwo}%
	\providecommand \href [0]{\begingroup \@sanitize@url \@href}%
	\providecommand \@href[1]{\@@startlink{#1}\@@href}%
	\providecommand \@@href[1]{\endgroup#1\@@endlink}%
	\providecommand \@sanitize@url [0]{\catcode `\\12\catcode `\$12\catcode `\&12\catcode `\#12\catcode `\^12\catcode `\_12\catcode `\%12\relax}%
	\providecommand \@@startlink[1]{}%
	\providecommand \@@endlink[0]{}%
	\providecommand \url  [0]{\begingroup\@sanitize@url \@url }%
	\providecommand \@url [1]{\endgroup\@href {#1}{\urlprefix }}%
	\providecommand \urlprefix  [0]{URL }%
	\providecommand \Eprint [0]{\href }%
	\providecommand \doibase [0]{https://doi.org/}%
	\providecommand \selectlanguage [0]{\@gobble}%
	\providecommand \bibinfo  [0]{\@secondoftwo}%
	\providecommand \bibfield  [0]{\@secondoftwo}%
	\providecommand \translation [1]{[#1]}%
	\providecommand \BibitemOpen [0]{}%
	\providecommand \bibitemStop [0]{}%
	\providecommand \bibitemNoStop [0]{.\EOS\space}%
	\providecommand \EOS [0]{\spacefactor3000\relax}%
	\providecommand \BibitemShut  [1]{\csname bibitem#1\endcsname}%
	\let\auto@bib@innerbib\@empty
	\bibitem [{\citenamefont {Rocha}\ \emph {et~al.}(2023)\citenamefont {Rocha}, \citenamefont {Wagner}, \citenamefont {Denicol}, \citenamefont {Noronha},\ and\ \citenamefont {Rischke}}]{Rocha:2023ilf}%
	\BibitemOpen
	\bibfield  {author} {\bibinfo {author} {\bibfnamefont {G.~S.}\ \bibnamefont {Rocha}}, \bibinfo {author} {\bibfnamefont {D.}~\bibnamefont {Wagner}}, \bibinfo {author} {\bibfnamefont {G.~S.}\ \bibnamefont {Denicol}}, \bibinfo {author} {\bibfnamefont {J.}~\bibnamefont {Noronha}},\ and\ \bibinfo {author} {\bibfnamefont {D.~H.}\ \bibnamefont {Rischke}},\ }\bibfield  {title} {\bibinfo {title} {{Theories of Relativistic Dissipative Fluid Dynamics}},\ }\href@noop {} {\  (\bibinfo {year} {2023})},\ \Eprint {https://arxiv.org/abs/2311.15063} {arXiv:2311.15063 [nucl-th]} \BibitemShut {NoStop}%
	\bibitem [{\citenamefont {Heinz}\ and\ \citenamefont {Snellings}(2013)}]{Heinz:2013th}%
	\BibitemOpen
	\bibfield  {author} {\bibinfo {author} {\bibfnamefont {U.}~\bibnamefont {Heinz}}\ and\ \bibinfo {author} {\bibfnamefont {R.}~\bibnamefont {Snellings}},\ }\bibfield  {title} {\bibinfo {title} {Collective flow and viscosity in relativistic heavy-ion collisions},\ }\href {https://doi.org/10.1146/annurev-nucl-102212-170540} {\bibfield  {journal} {\bibinfo  {journal} {Ann. Rev. Nucl. Part. Sci.}\ }\textbf {\bibinfo {volume} {63}},\ \bibinfo {pages} {123} (\bibinfo {year} {2013})},\ \Eprint {https://arxiv.org/abs/1301.2826} {arXiv:1301.2826 [nucl-th]} \BibitemShut {NoStop}%
	\bibitem [{\citenamefont {Gale}\ \emph {et~al.}(2013)\citenamefont {Gale}, \citenamefont {Jeon},\ and\ \citenamefont {Schenke}}]{Gale:2013da}%
	\BibitemOpen
	\bibfield  {author} {\bibinfo {author} {\bibfnamefont {C.}~\bibnamefont {Gale}}, \bibinfo {author} {\bibfnamefont {S.}~\bibnamefont {Jeon}},\ and\ \bibinfo {author} {\bibfnamefont {B.}~\bibnamefont {Schenke}},\ }\bibfield  {title} {\bibinfo {title} {{Hydrodynamic Modeling of Heavy-Ion Collisions}},\ }\href {https://doi.org/10.1142/S0217751X13400113} {\bibfield  {journal} {\bibinfo  {journal} {Int. J. Mod. Phys. A}\ }\textbf {\bibinfo {volume} {28}},\ \bibinfo {pages} {1340011} (\bibinfo {year} {2013})},\ \Eprint {https://arxiv.org/abs/1301.5893} {arXiv:1301.5893 [nucl-th]} \BibitemShut {NoStop}%
	\bibitem [{\citenamefont {Romatschke}\ and\ \citenamefont {Romatschke}(2019)}]{Romatschke:2017ejr}%
	\BibitemOpen
	\bibfield  {author} {\bibinfo {author} {\bibfnamefont {P.}~\bibnamefont {Romatschke}}\ and\ \bibinfo {author} {\bibfnamefont {U.}~\bibnamefont {Romatschke}},\ }\href@noop {} {\emph {\bibinfo {title} {Relativistic Fluid Dynamics In and Out of Equilibrium}}},\ Cambridge Monographs on Mathematical Physics\ (\bibinfo  {publisher} {Cambridge University Press},\ \bibinfo {year} {2019})\ \Eprint {https://arxiv.org/abs/1712.05815} {arXiv:1712.05815 [nucl-th]} \BibitemShut {NoStop}%
	\bibitem [{\citenamefont {Baiotti}\ and\ \citenamefont {Rezzolla}(2017)}]{Baiotti:2016qnr}%
	\BibitemOpen
	\bibfield  {author} {\bibinfo {author} {\bibfnamefont {L.}~\bibnamefont {Baiotti}}\ and\ \bibinfo {author} {\bibfnamefont {L.}~\bibnamefont {Rezzolla}},\ }\bibfield  {title} {\bibinfo {title} {Binary neutron star mergers: a review of einstein's richest laboratory},\ }\href {https://doi.org/10.1088/1361-6633/aa67bb} {\bibfield  {journal} {\bibinfo  {journal} {Rept. Prog. Phys.}\ }\textbf {\bibinfo {volume} {80}},\ \bibinfo {pages} {096901} (\bibinfo {year} {2017})},\ \Eprint {https://arxiv.org/abs/1607.03540} {arXiv:1607.03540 [gr-qc]} \BibitemShut {NoStop}%
	\bibitem [{\citenamefont {Alford}\ \emph {et~al.}(2018)\citenamefont {Alford}, \citenamefont {Bovard}, \citenamefont {Hanauske}, \citenamefont {Rezzolla},\ and\ \citenamefont {Schwenzer}}]{Alford:2017rxf}%
	\BibitemOpen
	\bibfield  {author} {\bibinfo {author} {\bibfnamefont {M.~G.}\ \bibnamefont {Alford}}, \bibinfo {author} {\bibfnamefont {L.}~\bibnamefont {Bovard}}, \bibinfo {author} {\bibfnamefont {M.}~\bibnamefont {Hanauske}}, \bibinfo {author} {\bibfnamefont {L.}~\bibnamefont {Rezzolla}},\ and\ \bibinfo {author} {\bibfnamefont {K.}~\bibnamefont {Schwenzer}},\ }\bibfield  {title} {\bibinfo {title} {Viscous dissipation and heat conduction in binary neutron-star mergers},\ }\href {https://doi.org/10.1103/PhysRevLett.120.041101} {\bibfield  {journal} {\bibinfo  {journal} {Phys. Rev. Lett.}\ }\textbf {\bibinfo {volume} {120}},\ \bibinfo {pages} {041101} (\bibinfo {year} {2018})},\ \Eprint {https://arxiv.org/abs/1707.09475} {arXiv:1707.09475 [gr-qc]} \BibitemShut {NoStop}%
	\bibitem [{\citenamefont {Most}\ \emph {et~al.}(2024)\citenamefont {Most}, \citenamefont {Haber}, \citenamefont {Harris}, \citenamefont {Zhang}, \citenamefont {Alford},\ and\ \citenamefont {Noronha}}]{Most:2022yhe}%
	\BibitemOpen
	\bibfield  {author} {\bibinfo {author} {\bibfnamefont {E.~R.}\ \bibnamefont {Most}}, \bibinfo {author} {\bibfnamefont {A.}~\bibnamefont {Haber}}, \bibinfo {author} {\bibfnamefont {S.~P.}\ \bibnamefont {Harris}}, \bibinfo {author} {\bibfnamefont {Z.}~\bibnamefont {Zhang}}, \bibinfo {author} {\bibfnamefont {M.~G.}\ \bibnamefont {Alford}},\ and\ \bibinfo {author} {\bibfnamefont {J.}~\bibnamefont {Noronha}},\ }\bibfield  {title} {\bibinfo {title} {{Emergence of Microphysical Bulk Viscosity in Binary Neutron Star Postmerger Dynamics}},\ }\href {https://doi.org/10.3847/2041-8213/ad454f} {\bibfield  {journal} {\bibinfo  {journal} {Astrophys. J. Lett.}\ }\textbf {\bibinfo {volume} {967}},\ \bibinfo {pages} {L14} (\bibinfo {year} {2024})},\ \Eprint {https://arxiv.org/abs/2207.00442} {arXiv:2207.00442 [astro-ph.HE]} \BibitemShut {NoStop}%
	\bibitem [{\citenamefont {Foucart}\ \emph {et~al.}(2017)\citenamefont {Foucart}, \citenamefont {Chandra}, \citenamefont {Gammie}, \citenamefont {Quataert},\ and\ \citenamefont {Tchekhovskoy}}]{Foucart:2017axc}%
	\BibitemOpen
	\bibfield  {author} {\bibinfo {author} {\bibfnamefont {F.}~\bibnamefont {Foucart}}, \bibinfo {author} {\bibfnamefont {M.}~\bibnamefont {Chandra}}, \bibinfo {author} {\bibfnamefont {C.~F.}\ \bibnamefont {Gammie}}, \bibinfo {author} {\bibfnamefont {E.}~\bibnamefont {Quataert}},\ and\ \bibinfo {author} {\bibfnamefont {A.}~\bibnamefont {Tchekhovskoy}},\ }\bibfield  {title} {\bibinfo {title} {{How important is non-ideal physics in simulations of sub-Eddington accretion on to spinning black holes?}},\ }\href {https://doi.org/10.1093/mnras/stx1368} {\bibfield  {journal} {\bibinfo  {journal} {Mon. Not. Roy. Astron. Soc.}\ }\textbf {\bibinfo {volume} {470}},\ \bibinfo {pages} {2240} (\bibinfo {year} {2017})},\ \Eprint {https://arxiv.org/abs/1706.01533} {arXiv:1706.01533 [astro-ph.HE]} \BibitemShut {NoStop}%
	\bibitem [{\citenamefont {Khachatryan}\ \emph {et~al.}(2015)\citenamefont {Khachatryan} \emph {et~al.}}]{CMS:2015yux}%
	\BibitemOpen
	\bibfield  {author} {\bibinfo {author} {\bibfnamefont {V.}~\bibnamefont {Khachatryan}} \emph {et~al.} (\bibinfo {collaboration} {CMS}),\ }\bibfield  {title} {\bibinfo {title} {{Evidence for Collective Multiparticle Correlations in p-Pb Collisions}},\ }\href {https://doi.org/10.1103/PhysRevLett.115.012301} {\bibfield  {journal} {\bibinfo  {journal} {Phys. Rev. Lett.}\ }\textbf {\bibinfo {volume} {115}},\ \bibinfo {pages} {012301} (\bibinfo {year} {2015})},\ \Eprint {https://arxiv.org/abs/1502.05382} {arXiv:1502.05382 [nucl-ex]} \BibitemShut {NoStop}%
	\bibitem [{\citenamefont {Aad}\ \emph {et~al.}(2016)\citenamefont {Aad} \emph {et~al.}}]{ATLAS:2015hzw}%
	\BibitemOpen
	\bibfield  {author} {\bibinfo {author} {\bibfnamefont {G.}~\bibnamefont {Aad}} \emph {et~al.} (\bibinfo {collaboration} {ATLAS}),\ }\bibfield  {title} {\bibinfo {title} {{Observation of Long-Range Elliptic Azimuthal Anisotropies in $\sqrt{s}=$13 and 2.76 TeV $pp$ Collisions with the ATLAS Detector}},\ }\href {https://doi.org/10.1103/PhysRevLett.116.172301} {\bibfield  {journal} {\bibinfo  {journal} {Phys. Rev. Lett.}\ }\textbf {\bibinfo {volume} {116}},\ \bibinfo {pages} {172301} (\bibinfo {year} {2016})},\ \Eprint {https://arxiv.org/abs/1509.04776} {arXiv:1509.04776 [hep-ex]} \BibitemShut {NoStop}%
	\bibitem [{\citenamefont {Khachatryan}\ \emph {et~al.}(2017)\citenamefont {Khachatryan} \emph {et~al.}}]{CMS:2016fnw}%
	\BibitemOpen
	\bibfield  {author} {\bibinfo {author} {\bibfnamefont {V.}~\bibnamefont {Khachatryan}} \emph {et~al.} (\bibinfo {collaboration} {CMS}),\ }\bibfield  {title} {\bibinfo {title} {{Evidence for collectivity in pp collisions at the LHC}},\ }\href {https://doi.org/10.1016/j.physletb.2016.12.009} {\bibfield  {journal} {\bibinfo  {journal} {Phys. Lett. B}\ }\textbf {\bibinfo {volume} {765}},\ \bibinfo {pages} {193} (\bibinfo {year} {2017})},\ \Eprint {https://arxiv.org/abs/1606.06198} {arXiv:1606.06198 [nucl-ex]} \BibitemShut {NoStop}%
	\bibitem [{\citenamefont {Weller}\ and\ \citenamefont {Romatschke}(2017)}]{Weller:2017tsr}%
	\BibitemOpen
	\bibfield  {author} {\bibinfo {author} {\bibfnamefont {R.~D.}\ \bibnamefont {Weller}}\ and\ \bibinfo {author} {\bibfnamefont {P.}~\bibnamefont {Romatschke}},\ }\bibfield  {title} {\bibinfo {title} {{One fluid to rule them all: viscous hydrodynamic description of event-by-event central p+p, p+Pb and Pb+Pb collisions at $\sqrt{s}=5.02$ TeV}},\ }\href {https://doi.org/10.1016/j.physletb.2017.09.077} {\bibfield  {journal} {\bibinfo  {journal} {Phys. Lett. B}\ }\textbf {\bibinfo {volume} {774}},\ \bibinfo {pages} {351} (\bibinfo {year} {2017})},\ \Eprint {https://arxiv.org/abs/1701.07145} {arXiv:1701.07145 [nucl-th]} \BibitemShut {NoStop}%
	\bibitem [{\citenamefont {Aidala}\ \emph {et~al.}(2019)\citenamefont {Aidala} \emph {et~al.}}]{PHENIX:2018lia}%
	\BibitemOpen
	\bibfield  {author} {\bibinfo {author} {\bibfnamefont {C.}~\bibnamefont {Aidala}} \emph {et~al.} (\bibinfo {collaboration} {PHENIX}),\ }\bibfield  {title} {\bibinfo {title} {{Creation of quark\textendash{}gluon plasma droplets with three distinct geometries}},\ }\href {https://doi.org/10.1038/s41567-018-0360-0} {\bibfield  {journal} {\bibinfo  {journal} {Nature Phys.}\ }\textbf {\bibinfo {volume} {15}},\ \bibinfo {pages} {214} (\bibinfo {year} {2019})},\ \Eprint {https://arxiv.org/abs/1805.02973} {arXiv:1805.02973 [nucl-ex]} \BibitemShut {NoStop}%
	\bibitem [{\citenamefont {Acharya}\ \emph {et~al.}(2019)\citenamefont {Acharya} \emph {et~al.}}]{ALICE:2019zfl}%
	\BibitemOpen
	\bibfield  {author} {\bibinfo {author} {\bibfnamefont {S.}~\bibnamefont {Acharya}} \emph {et~al.} (\bibinfo {collaboration} {ALICE}),\ }\bibfield  {title} {\bibinfo {title} {{Investigations of Anisotropic Flow Using Multiparticle Azimuthal Correlations in pp, p-Pb, Xe-Xe, and Pb-Pb Collisions at the LHC}},\ }\href {https://doi.org/10.1103/PhysRevLett.123.142301} {\bibfield  {journal} {\bibinfo  {journal} {Phys. Rev. Lett.}\ }\textbf {\bibinfo {volume} {123}},\ \bibinfo {pages} {142301} (\bibinfo {year} {2019})},\ \Eprint {https://arxiv.org/abs/1903.01790} {arXiv:1903.01790 [nucl-ex]} \BibitemShut {NoStop}%
	\bibitem [{\citenamefont {Noronha}\ \emph {et~al.}(2024)\citenamefont {Noronha}, \citenamefont {Schenke}, \citenamefont {Shen},\ and\ \citenamefont {Zhao}}]{Noronha:2024dtq}%
	\BibitemOpen
	\bibfield  {author} {\bibinfo {author} {\bibfnamefont {J.}~\bibnamefont {Noronha}}, \bibinfo {author} {\bibfnamefont {B.}~\bibnamefont {Schenke}}, \bibinfo {author} {\bibfnamefont {C.}~\bibnamefont {Shen}},\ and\ \bibinfo {author} {\bibfnamefont {W.}~\bibnamefont {Zhao}},\ }\bibfield  {title} {\bibinfo {title} {{Progress and challenges in small systems}},\ }\href {https://doi.org/10.1142/9789811294679_0004} {\bibfield  {journal} {\bibinfo  {journal} {Int. J. Mod. Phys. E}\ }\textbf {\bibinfo {volume} {33}},\ \bibinfo {pages} {2430005} (\bibinfo {year} {2024})},\ \Eprint {https://arxiv.org/abs/2401.09208} {arXiv:2401.09208 [nucl-th]} \BibitemShut {NoStop}%
	\bibitem [{\citenamefont {Grosse-Oetringhaus}\ and\ \citenamefont {Wiedemann}(2024)}]{Grosse-Oetringhaus:2024bwr}%
	\BibitemOpen
	\bibfield  {author} {\bibinfo {author} {\bibfnamefont {J.~F.}\ \bibnamefont {Grosse-Oetringhaus}}\ and\ \bibinfo {author} {\bibfnamefont {U.~A.}\ \bibnamefont {Wiedemann}},\ }\bibfield  {title} {\bibinfo {title} {{A Decade of Collectivity in Small Systems}},\ }\href@noop {} {\  (\bibinfo {year} {2024})},\ \Eprint {https://arxiv.org/abs/2407.07484} {arXiv:2407.07484 [hep-ex]} \BibitemShut {NoStop}%
	\bibitem [{\citenamefont {Chandra}\ \emph {et~al.}(2015)\citenamefont {Chandra}, \citenamefont {Gammie}, \citenamefont {Foucart},\ and\ \citenamefont {Quataert}}]{Chandra:2015iza}%
	\BibitemOpen
	\bibfield  {author} {\bibinfo {author} {\bibfnamefont {M.}~\bibnamefont {Chandra}}, \bibinfo {author} {\bibfnamefont {C.~F.}\ \bibnamefont {Gammie}}, \bibinfo {author} {\bibfnamefont {F.}~\bibnamefont {Foucart}},\ and\ \bibinfo {author} {\bibfnamefont {E.}~\bibnamefont {Quataert}},\ }\bibfield  {title} {\bibinfo {title} {{An Extended Magnetohydrodynamics Model for Relativistic Weakly Collisional Plasmas}},\ }\href {https://doi.org/10.1088/0004-637X/810/2/162} {\bibfield  {journal} {\bibinfo  {journal} {Astrophys. J.}\ }\textbf {\bibinfo {volume} {810}},\ \bibinfo {pages} {162} (\bibinfo {year} {2015})},\ \Eprint {https://arxiv.org/abs/1508.00878} {arXiv:1508.00878 [astro-ph.HE]} \BibitemShut {NoStop}%
	\bibitem [{\citenamefont {Akiyama}\ \emph {et~al.}(2022)\citenamefont {Akiyama} \emph {et~al.}}]{EventHorizonTelescope:2022wkp}%
	\BibitemOpen
	\bibfield  {author} {\bibinfo {author} {\bibfnamefont {K.}~\bibnamefont {Akiyama}} \emph {et~al.} (\bibinfo {collaboration} {Event Horizon Telescope}),\ }\bibfield  {title} {\bibinfo {title} {{First Sagittarius A* Event Horizon Telescope Results. I. The Shadow of the Supermassive Black Hole in the Center of the Milky Way}},\ }\href {https://doi.org/10.3847/2041-8213/ac6674} {\bibfield  {journal} {\bibinfo  {journal} {Astrophys. J. Lett.}\ }\textbf {\bibinfo {volume} {930}},\ \bibinfo {pages} {L12} (\bibinfo {year} {2022})},\ \Eprint {https://arxiv.org/abs/2311.08680} {arXiv:2311.08680 [astro-ph.HE]} \BibitemShut {NoStop}%
	\bibitem [{\citenamefont {Cordeiro}\ \emph {et~al.}(2024)\citenamefont {Cordeiro}, \citenamefont {Speranza}, \citenamefont {Ingles}, \citenamefont {Bemfica},\ and\ \citenamefont {Noronha}}]{Cordeiro:2023ljz}%
	\BibitemOpen
	\bibfield  {author} {\bibinfo {author} {\bibfnamefont {I.}~\bibnamefont {Cordeiro}}, \bibinfo {author} {\bibfnamefont {E.}~\bibnamefont {Speranza}}, \bibinfo {author} {\bibfnamefont {K.}~\bibnamefont {Ingles}}, \bibinfo {author} {\bibfnamefont {F.~S.}\ \bibnamefont {Bemfica}},\ and\ \bibinfo {author} {\bibfnamefont {J.}~\bibnamefont {Noronha}},\ }\bibfield  {title} {\bibinfo {title} {{Causality Bounds on Dissipative General-Relativistic Magnetohydrodynamics}},\ }\href {https://doi.org/10.1103/PhysRevLett.133.091401} {\bibfield  {journal} {\bibinfo  {journal} {Phys. Rev. Lett.}\ }\textbf {\bibinfo {volume} {133}},\ \bibinfo {pages} {091401} (\bibinfo {year} {2024})},\ \Eprint {https://arxiv.org/abs/2312.09970} {arXiv:2312.09970 [astro-ph.HE]} \BibitemShut {NoStop}%
	\bibitem [{\citenamefont {Abbasi}\ \emph {et~al.}(2025)\citenamefont {Abbasi}, \citenamefont {Kaminski},\ and\ \citenamefont {Rischke}}]{abbasi2025}%
	\BibitemOpen
	\bibfield  {author} {\bibinfo {author} {\bibfnamefont {N.}~\bibnamefont {Abbasi}}, \bibinfo {author} {\bibfnamefont {M.}~\bibnamefont {Kaminski}},\ and\ \bibinfo {author} {\bibfnamefont {D.~H.}\ \bibnamefont {Rischke}},\ }\href {https://arxiv.org/abs/2506.20500} {\bibinfo {title} {Comparison between causal and acausal diffusion: a schwinger-keldysh effective field theory perspective}} (\bibinfo {year} {2025}),\ \Eprint {https://arxiv.org/abs/2506.20500} {arXiv:2506.20500 [hep-th]} \BibitemShut {NoStop}%
	\bibitem [{\citenamefont {Eckart}(1940)}]{EckartViscous}%
	\BibitemOpen
	\bibfield  {author} {\bibinfo {author} {\bibfnamefont {C.}~\bibnamefont {Eckart}},\ }\bibfield  {title} {\bibinfo {title} {The thermodynamics of irreversible processes {III}. {R}elativistic theory of the simple fluid},\ }\href@noop {} {\bibfield  {journal} {\bibinfo  {journal} {Physical Review}\ }\textbf {\bibinfo {volume} {58}},\ \bibinfo {pages} {919} (\bibinfo {year} {1940})}\BibitemShut {NoStop}%
	\bibitem [{\citenamefont {Landau}\ and\ \citenamefont {Lifshitz}(1987)}]{LandauLifshitzFluids}%
	\BibitemOpen
	\bibfield  {author} {\bibinfo {author} {\bibfnamefont {L.~D.}\ \bibnamefont {Landau}}\ and\ \bibinfo {author} {\bibfnamefont {E.~M.}\ \bibnamefont {Lifshitz}},\ }\href@noop {} {\emph {\bibinfo {title} {Fluid Mechanics - Volume 6 (Corse of Theoretical Physics)}}},\ \bibinfo {edition} {2nd}\ ed.\ (\bibinfo  {publisher} {Butterworth-Heinemann},\ \bibinfo {year} {1987})\ p.\ \bibinfo {pages} {552}\BibitemShut {NoStop}%
	\bibitem [{\citenamefont {Hiscock}\ and\ \citenamefont {Lindblom}(1983)}]{Hiscock_Lindblom_stability_1983}%
	\BibitemOpen
	\bibfield  {author} {\bibinfo {author} {\bibfnamefont {W.~A.}\ \bibnamefont {Hiscock}}\ and\ \bibinfo {author} {\bibfnamefont {L.}~\bibnamefont {Lindblom}},\ }\bibfield  {title} {\bibinfo {title} {Stability and causality in dissipative relativistic fluids},\ }\href@noop {} {\bibfield  {journal} {\bibinfo  {journal} {Annals of Physics}\ }\textbf {\bibinfo {volume} {151}},\ \bibinfo {pages} {466} (\bibinfo {year} {1983})}\BibitemShut {NoStop}%
	\bibitem [{\citenamefont {Hiscock}\ and\ \citenamefont {Lindblom}(1985)}]{Hiscock_Lindblom_instability_1985}%
	\BibitemOpen
	\bibfield  {author} {\bibinfo {author} {\bibfnamefont {W.~A.}\ \bibnamefont {Hiscock}}\ and\ \bibinfo {author} {\bibfnamefont {L.}~\bibnamefont {Lindblom}},\ }\bibfield  {title} {\bibinfo {title} {Generic instabilities in first-order dissipative fluid theories},\ }\href@noop {} {\bibfield  {journal} {\bibinfo  {journal} {Phys. Rev. D}\ }\textbf {\bibinfo {volume} {31}},\ \bibinfo {pages} {725} (\bibinfo {year} {1985})}\BibitemShut {NoStop}%
	\bibitem [{\citenamefont {Hiscock}\ and\ \citenamefont {Lindblom}(1987)}]{Hiscock_Lindblom_acausality_1987}%
	\BibitemOpen
	\bibfield  {author} {\bibinfo {author} {\bibfnamefont {W.~A.}\ \bibnamefont {Hiscock}}\ and\ \bibinfo {author} {\bibfnamefont {L.}~\bibnamefont {Lindblom}},\ }\bibfield  {title} {\bibinfo {title} {Linear plane waves in dissipative relativistic fluids},\ }\href {https://doi.org/10.1103/PhysRevD.35.3723} {\bibfield  {journal} {\bibinfo  {journal} {Phys. Rev. D}\ }\textbf {\bibinfo {volume} {35}},\ \bibinfo {pages} {3723} (\bibinfo {year} {1987})}\BibitemShut {NoStop}%
	\bibitem [{\citenamefont {Israel}(1976{\natexlab{a}})}]{MIS-2}%
	\BibitemOpen
	\bibfield  {author} {\bibinfo {author} {\bibfnamefont {W.}~\bibnamefont {Israel}},\ }\bibfield  {title} {\bibinfo {title} {Nonstationary irreversible thermodynamics: A causal relativistic theory},\ }\href@noop {} {\bibfield  {journal} {\bibinfo  {journal} {Ann. Phys.}\ }\textbf {\bibinfo {volume} {100}},\ \bibinfo {pages} {310} (\bibinfo {year} {1976}{\natexlab{a}})}\BibitemShut {NoStop}%
	\bibitem [{\citenamefont {Israel}\ and\ \citenamefont {Stewart}(1979{\natexlab{a}})}]{MIS-6}%
	\BibitemOpen
	\bibfield  {author} {\bibinfo {author} {\bibfnamefont {W.}~\bibnamefont {Israel}}\ and\ \bibinfo {author} {\bibfnamefont {J.~M.}\ \bibnamefont {Stewart}},\ }\bibfield  {title} {\bibinfo {title} {Transient relativistic thermodynamics and kinetic theory},\ }\href@noop {} {\bibfield  {journal} {\bibinfo  {journal} {Ann. Phys.}\ }\textbf {\bibinfo {volume} {118}},\ \bibinfo {pages} {341} (\bibinfo {year} {1979}{\natexlab{a}})}\BibitemShut {NoStop}%
	\bibitem [{\citenamefont {Bemfica}\ \emph {et~al.}(2018)\citenamefont {Bemfica}, \citenamefont {Disconzi},\ and\ \citenamefont {Noronha}}]{Bemfica:2017wps}%
	\BibitemOpen
	\bibfield  {author} {\bibinfo {author} {\bibfnamefont {F.~S.}\ \bibnamefont {Bemfica}}, \bibinfo {author} {\bibfnamefont {M.~M.}\ \bibnamefont {Disconzi}},\ and\ \bibinfo {author} {\bibfnamefont {J.}~\bibnamefont {Noronha}},\ }\bibfield  {title} {\bibinfo {title} {{Causality and existence of solutions of relativistic viscous fluid dynamics with gravity}},\ }\href {https://doi.org/10.1103/PhysRevD.98.104064} {\bibfield  {journal} {\bibinfo  {journal} {Phys. Rev. D}\ }\textbf {\bibinfo {volume} {98}},\ \bibinfo {pages} {104064} (\bibinfo {year} {2018})},\ \Eprint {https://arxiv.org/abs/1708.06255} {arXiv:1708.06255 [gr-qc]} \BibitemShut {NoStop}%
	\bibitem [{\citenamefont {Kovtun}(2019)}]{Kovtun:2019hdm}%
	\BibitemOpen
	\bibfield  {author} {\bibinfo {author} {\bibfnamefont {P.}~\bibnamefont {Kovtun}},\ }\bibfield  {title} {\bibinfo {title} {{First-order relativistic hydrodynamics is stable}},\ }\href {https://doi.org/10.1007/JHEP10(2019)034} {\bibfield  {journal} {\bibinfo  {journal} {JHEP}\ }\textbf {\bibinfo {volume} {10}},\ \bibinfo {pages} {034}},\ \Eprint {https://arxiv.org/abs/1907.08191} {arXiv:1907.08191 [hep-th]} \BibitemShut {NoStop}%
	\bibitem [{\citenamefont {Bemfica}\ \emph {et~al.}(2019{\natexlab{a}})\citenamefont {Bemfica}, \citenamefont {Disconzi},\ and\ \citenamefont {Noronha}}]{Bemfica:2019knx}%
	\BibitemOpen
	\bibfield  {author} {\bibinfo {author} {\bibfnamefont {F.~S.}\ \bibnamefont {Bemfica}}, \bibinfo {author} {\bibfnamefont {M.~M.}\ \bibnamefont {Disconzi}},\ and\ \bibinfo {author} {\bibfnamefont {J.}~\bibnamefont {Noronha}},\ }\bibfield  {title} {\bibinfo {title} {{Nonlinear Causality of General First-Order Relativistic Viscous Hydrodynamics}},\ }\href {https://doi.org/10.1103/PhysRevD.100.104020} {\bibfield  {journal} {\bibinfo  {journal} {Phys. Rev. D}\ }\textbf {\bibinfo {volume} {100}},\ \bibinfo {pages} {104020} (\bibinfo {year} {2019}{\natexlab{a}})},\ \Eprint {https://arxiv.org/abs/1907.12695} {arXiv:1907.12695 [gr-qc]} \BibitemShut {NoStop}%
	\bibitem [{\citenamefont {Hoult}\ and\ \citenamefont {Kovtun}(2020)}]{Hoult:2020eho}%
	\BibitemOpen
	\bibfield  {author} {\bibinfo {author} {\bibfnamefont {R.~E.}\ \bibnamefont {Hoult}}\ and\ \bibinfo {author} {\bibfnamefont {P.}~\bibnamefont {Kovtun}},\ }\bibfield  {title} {\bibinfo {title} {{Stable and causal relativistic Navier-Stokes equations}},\ }\href {https://doi.org/10.1007/JHEP06(2020)067} {\bibfield  {journal} {\bibinfo  {journal} {JHEP}\ }\textbf {\bibinfo {volume} {06}},\ \bibinfo {pages} {067}},\ \Eprint {https://arxiv.org/abs/2004.04102} {arXiv:2004.04102 [hep-th]} \BibitemShut {NoStop}%
	\bibitem [{\citenamefont {Bemfica}\ \emph {et~al.}(2022)\citenamefont {Bemfica}, \citenamefont {Disconzi},\ and\ \citenamefont {Noronha}}]{Bemfica:2020zjp}%
	\BibitemOpen
	\bibfield  {author} {\bibinfo {author} {\bibfnamefont {F.~S.}\ \bibnamefont {Bemfica}}, \bibinfo {author} {\bibfnamefont {M.~M.}\ \bibnamefont {Disconzi}},\ and\ \bibinfo {author} {\bibfnamefont {J.}~\bibnamefont {Noronha}},\ }\bibfield  {title} {\bibinfo {title} {First-order general-relativistic viscous fluid dynamics},\ }\href {https://doi.org/10.1103/PhysRevX.12.021044} {\bibfield  {journal} {\bibinfo  {journal} {Phys. Rev. X}\ }\textbf {\bibinfo {volume} {12}},\ \bibinfo {pages} {021044} (\bibinfo {year} {2022})}\BibitemShut {NoStop}%
	\bibitem [{\citenamefont {Abboud}\ \emph {et~al.}(2024)\citenamefont {Abboud}, \citenamefont {Speranza},\ and\ \citenamefont {Noronha}}]{Abboud:2023hos}%
	\BibitemOpen
	\bibfield  {author} {\bibinfo {author} {\bibfnamefont {N.}~\bibnamefont {Abboud}}, \bibinfo {author} {\bibfnamefont {E.}~\bibnamefont {Speranza}},\ and\ \bibinfo {author} {\bibfnamefont {J.}~\bibnamefont {Noronha}},\ }\bibfield  {title} {\bibinfo {title} {{Causal and stable first-order chiral hydrodynamics}},\ }\href {https://doi.org/10.1103/PhysRevD.109.094007} {\bibfield  {journal} {\bibinfo  {journal} {Phys. Rev. D}\ }\textbf {\bibinfo {volume} {109}},\ \bibinfo {pages} {094007} (\bibinfo {year} {2024})},\ \Eprint {https://arxiv.org/abs/2308.02928} {arXiv:2308.02928 [hep-th]} \BibitemShut {NoStop}%
	\bibitem [{\citenamefont {Baier}\ \emph {et~al.}(2008)\citenamefont {Baier}, \citenamefont {Romatschke}, \citenamefont {Son}, \citenamefont {Starinets},\ and\ \citenamefont {Stephanov}}]{Baier:2007ix}%
	\BibitemOpen
	\bibfield  {author} {\bibinfo {author} {\bibfnamefont {R.}~\bibnamefont {Baier}}, \bibinfo {author} {\bibfnamefont {P.}~\bibnamefont {Romatschke}}, \bibinfo {author} {\bibfnamefont {D.~T.}\ \bibnamefont {Son}}, \bibinfo {author} {\bibfnamefont {A.~O.}\ \bibnamefont {Starinets}},\ and\ \bibinfo {author} {\bibfnamefont {M.~A.}\ \bibnamefont {Stephanov}},\ }\bibfield  {title} {\bibinfo {title} {Relativistic viscous hydrodynamics, conformal invariance, and holography},\ }\href {https://doi.org/10.1088/1126-6708/2008/04/100} {\bibfield  {journal} {\bibinfo  {journal} {JHEP}\ }\textbf {\bibinfo {volume} {04}},\ \bibinfo {pages} {100}},\ \Eprint {https://arxiv.org/abs/0712.2451} {arXiv:0712.2451 [hep-th]} \BibitemShut {NoStop}%
	\bibitem [{\citenamefont {Denicol}\ \emph {et~al.}(2012)\citenamefont {Denicol}, \citenamefont {Niemi}, \citenamefont {Molnar},\ and\ \citenamefont {Rischke}}]{Denicol:2012cn}%
	\BibitemOpen
	\bibfield  {author} {\bibinfo {author} {\bibfnamefont {G.~S.}\ \bibnamefont {Denicol}}, \bibinfo {author} {\bibfnamefont {H.}~\bibnamefont {Niemi}}, \bibinfo {author} {\bibfnamefont {E.}~\bibnamefont {Molnar}},\ and\ \bibinfo {author} {\bibfnamefont {D.~H.}\ \bibnamefont {Rischke}},\ }\bibfield  {title} {\bibinfo {title} {Derivation of transient relativistic fluid dynamics from the {B}oltzmann equation},\ }\href {https://doi.org/10.1103/PhysRevD.85.114047, 10.1103/PhysRevD.91.039902} {\bibfield  {journal} {\bibinfo  {journal} {Phys. Rev.}\ }\textbf {\bibinfo {volume} {D85}},\ \bibinfo {pages} {114047} (\bibinfo {year} {2012})},\ \bibinfo {note} {[Erratum: Phys. Rev.D91,no.3,039902(2015)]},\ \Eprint {https://arxiv.org/abs/1202.4551} {arXiv:1202.4551 [nucl-th]} \BibitemShut {NoStop}%
	\bibitem [{\citenamefont {Noronha}\ \emph {et~al.}(2022)\citenamefont {Noronha}, \citenamefont {Spali\'nski},\ and\ \citenamefont {Speranza}}]{Noronha:2021syv}%
	\BibitemOpen
	\bibfield  {author} {\bibinfo {author} {\bibfnamefont {J.}~\bibnamefont {Noronha}}, \bibinfo {author} {\bibfnamefont {M.}~\bibnamefont {Spali\'nski}},\ and\ \bibinfo {author} {\bibfnamefont {E.}~\bibnamefont {Speranza}},\ }\bibfield  {title} {\bibinfo {title} {{Transient Relativistic Fluid Dynamics in a General Hydrodynamic Frame}},\ }\href {https://doi.org/10.1103/PhysRevLett.128.252302} {\bibfield  {journal} {\bibinfo  {journal} {Phys. Rev. Lett.}\ }\textbf {\bibinfo {volume} {128}},\ \bibinfo {pages} {252302} (\bibinfo {year} {2022})},\ \Eprint {https://arxiv.org/abs/2105.01034} {arXiv:2105.01034 [nucl-th]} \BibitemShut {NoStop}%
	\bibitem [{\citenamefont {Brito}\ and\ \citenamefont {Denicol}(2020)}]{Brito:2020nou}%
	\BibitemOpen
	\bibfield  {author} {\bibinfo {author} {\bibfnamefont {C.}~\bibnamefont {Brito}}\ and\ \bibinfo {author} {\bibfnamefont {G.}~\bibnamefont {Denicol}},\ }\bibfield  {title} {\bibinfo {title} {{Linear stability of Israel-Stewart theory in the presence of net-charge diffusion}},\ }\href@noop {} {\  (\bibinfo {year} {2020})},\ \Eprint {https://arxiv.org/abs/2007.16141} {arXiv:2007.16141 [nucl-th]} \BibitemShut {NoStop}%
	\bibitem [{\citenamefont {de~Brito}\ \emph {et~al.}(2025)\citenamefont {de~Brito}, \citenamefont {Kushwah},\ and\ \citenamefont {Denicol}}]{deBrito:2025jaz}%
	\BibitemOpen
	\bibfield  {author} {\bibinfo {author} {\bibfnamefont {C.~V.~P.}\ \bibnamefont {de~Brito}}, \bibinfo {author} {\bibfnamefont {K.}~\bibnamefont {Kushwah}},\ and\ \bibinfo {author} {\bibfnamefont {G.~S.}\ \bibnamefont {Denicol}},\ }\bibfield  {title} {\bibinfo {title} {{Causality and stability of magnetohydrodynamics for an ultrarelativistic locally neutral two-component gas}},\ }\href@noop {} {\  (\bibinfo {year} {2025})},\ \Eprint {https://arxiv.org/abs/2505.10397} {arXiv:2505.10397 [nucl-th]} \BibitemShut {NoStop}%
	\bibitem [{\citenamefont {Floerchinger}\ and\ \citenamefont {Grossi}(2018)}]{Floerchinger_2018}%
	\BibitemOpen
	\bibfield  {author} {\bibinfo {author} {\bibfnamefont {S.}~\bibnamefont {Floerchinger}}\ and\ \bibinfo {author} {\bibfnamefont {E.}~\bibnamefont {Grossi}},\ }\bibfield  {title} {\bibinfo {title} {Causality of fluid dynamics for high-energy nuclear collisions},\ }\bibfield  {journal} {\bibinfo  {journal} {Journal of High Energy Physics}\ }\textbf {\bibinfo {volume} {2018}},\ \href {https://doi.org/10.1007/jhep08(2018)186} {10.1007/jhep08(2018)186} (\bibinfo {year} {2018})\BibitemShut {NoStop}%
	\bibitem [{\citenamefont {Bemfica}\ \emph {et~al.}(2019{\natexlab{b}})\citenamefont {Bemfica}, \citenamefont {Disconzi},\ and\ \citenamefont {Noronha}}]{Bemfica:2019cop}%
	\BibitemOpen
	\bibfield  {author} {\bibinfo {author} {\bibfnamefont {F.~S.}\ \bibnamefont {Bemfica}}, \bibinfo {author} {\bibfnamefont {M.~M.}\ \bibnamefont {Disconzi}},\ and\ \bibinfo {author} {\bibfnamefont {J.}~\bibnamefont {Noronha}},\ }\bibfield  {title} {\bibinfo {title} {{Causality of the Einstein-Israel-Stewart Theory with Bulk Viscosity}},\ }\href {https://doi.org/10.1103/PhysRevLett.122.221602} {\bibfield  {journal} {\bibinfo  {journal} {Phys. Rev. Lett.}\ }\textbf {\bibinfo {volume} {122}},\ \bibinfo {pages} {221602} (\bibinfo {year} {2019}{\natexlab{b}})},\ \Eprint {https://arxiv.org/abs/1901.06701} {arXiv:1901.06701 [gr-qc]} \BibitemShut {NoStop}%
	\bibitem [{\citenamefont {Disconzi}\ \emph {et~al.}(2020)\citenamefont {Disconzi}, \citenamefont {Hoang},\ and\ \citenamefont {Radosz}}]{Disconzi:2020ijk}%
	\BibitemOpen
	\bibfield  {author} {\bibinfo {author} {\bibfnamefont {M.~M.}\ \bibnamefont {Disconzi}}, \bibinfo {author} {\bibfnamefont {V.}~\bibnamefont {Hoang}},\ and\ \bibinfo {author} {\bibfnamefont {M.}~\bibnamefont {Radosz}},\ }\bibfield  {title} {\bibinfo {title} {{Breakdown of smooth solutions to the M\"uller-Israel-Stewart equations of relativistic viscous fluids}},\ }\href@noop {} {\  (\bibinfo {year} {2020})},\ \Eprint {https://arxiv.org/abs/2008.03841} {arXiv:2008.03841 [math.AP]} \BibitemShut {NoStop}%
	\bibitem [{\citenamefont {Bemfica}\ \emph {et~al.}(2020)\citenamefont {Bemfica}, \citenamefont {Disconzi}, \citenamefont {Hoang}, \citenamefont {Noronha},\ and\ \citenamefont {Radosz}}]{Bemfica:2020xym}%
	\BibitemOpen
	\bibfield  {author} {\bibinfo {author} {\bibfnamefont {F.~S.}\ \bibnamefont {Bemfica}}, \bibinfo {author} {\bibfnamefont {M.~M.}\ \bibnamefont {Disconzi}}, \bibinfo {author} {\bibfnamefont {V.}~\bibnamefont {Hoang}}, \bibinfo {author} {\bibfnamefont {J.}~\bibnamefont {Noronha}},\ and\ \bibinfo {author} {\bibfnamefont {M.}~\bibnamefont {Radosz}},\ }\bibfield  {title} {\bibinfo {title} {{Nonlinear Constraints on Relativistic Fluids Far From Equilibrium}},\ }\href@noop {} {\  (\bibinfo {year} {2020})},\ \Eprint {https://arxiv.org/abs/2005.11632} {arXiv:2005.11632 [hep-th]} \BibitemShut {NoStop}%
	\bibitem [{\citenamefont {Plumberg}\ \emph {et~al.}(2022)\citenamefont {Plumberg}, \citenamefont {Almaalol}, \citenamefont {Dore}, \citenamefont {Noronha},\ and\ \citenamefont {Noronha-Hostler}}]{Plumberg:2021bme}%
	\BibitemOpen
	\bibfield  {author} {\bibinfo {author} {\bibfnamefont {C.}~\bibnamefont {Plumberg}}, \bibinfo {author} {\bibfnamefont {D.}~\bibnamefont {Almaalol}}, \bibinfo {author} {\bibfnamefont {T.}~\bibnamefont {Dore}}, \bibinfo {author} {\bibfnamefont {J.}~\bibnamefont {Noronha}},\ and\ \bibinfo {author} {\bibfnamefont {J.}~\bibnamefont {Noronha-Hostler}},\ }\bibfield  {title} {\bibinfo {title} {{Causality violations in realistic simulations of heavy-ion collisions}},\ }\href {https://doi.org/10.1103/PhysRevC.105.L061901} {\bibfield  {journal} {\bibinfo  {journal} {Phys. Rev. C}\ }\textbf {\bibinfo {volume} {105}},\ \bibinfo {pages} {L061901} (\bibinfo {year} {2022})},\ \Eprint {https://arxiv.org/abs/2103.15889} {arXiv:2103.15889 [nucl-th]} \BibitemShut {NoStop}%
	\bibitem [{\citenamefont {Chiu}\ and\ \citenamefont {Shen}(2021)}]{Chiu:2021muk}%
	\BibitemOpen
	\bibfield  {author} {\bibinfo {author} {\bibfnamefont {C.}~\bibnamefont {Chiu}}\ and\ \bibinfo {author} {\bibfnamefont {C.}~\bibnamefont {Shen}},\ }\bibfield  {title} {\bibinfo {title} {{Exploring theoretical uncertainties in the hydrodynamic description of relativistic heavy-ion collisions}},\ }\href {https://doi.org/10.1103/PhysRevC.103.064901} {\bibfield  {journal} {\bibinfo  {journal} {Phys. Rev. C}\ }\textbf {\bibinfo {volume} {103}},\ \bibinfo {pages} {064901} (\bibinfo {year} {2021})},\ \Eprint {https://arxiv.org/abs/2103.09848} {arXiv:2103.09848 [nucl-th]} \BibitemShut {NoStop}%
	\bibitem [{\citenamefont {Krupczak}\ \emph {et~al.}(2023)\citenamefont {Krupczak} \emph {et~al.}}]{Krupczak:2023jpa}%
	\BibitemOpen
	\bibfield  {author} {\bibinfo {author} {\bibfnamefont {R.}~\bibnamefont {Krupczak}} \emph {et~al.},\ }\bibfield  {title} {\bibinfo {title} {{Causality violations in simulations of large and small heavy-ion collisions}},\ }\href@noop {} {\  (\bibinfo {year} {2023})},\ \Eprint {https://arxiv.org/abs/2311.02210} {arXiv:2311.02210 [nucl-th]} \BibitemShut {NoStop}%
	\bibitem [{\citenamefont {Domingues}\ \emph {et~al.}(2024)\citenamefont {Domingues}, \citenamefont {Krupczak}, \citenamefont {Noronha}, \citenamefont {da~Silva}, \citenamefont {Paquet},\ and\ \citenamefont {Luzum}}]{Domingues:2024pom}%
	\BibitemOpen
	\bibfield  {author} {\bibinfo {author} {\bibfnamefont {T.~S.}\ \bibnamefont {Domingues}}, \bibinfo {author} {\bibfnamefont {R.}~\bibnamefont {Krupczak}}, \bibinfo {author} {\bibfnamefont {J.}~\bibnamefont {Noronha}}, \bibinfo {author} {\bibfnamefont {T.~N.}\ \bibnamefont {da~Silva}}, \bibinfo {author} {\bibfnamefont {J.-F.}\ \bibnamefont {Paquet}},\ and\ \bibinfo {author} {\bibfnamefont {M.}~\bibnamefont {Luzum}},\ }\bibfield  {title} {\bibinfo {title} {{Effect of causality constraints on Bayesian analyses of heavy-ion collisions}},\ }\href {https://doi.org/10.1103/PhysRevC.110.064904} {\bibfield  {journal} {\bibinfo  {journal} {Phys. Rev. C}\ }\textbf {\bibinfo {volume} {110}},\ \bibinfo {pages} {064904} (\bibinfo {year} {2024})},\ \Eprint {https://arxiv.org/abs/2409.17127} {arXiv:2409.17127 [nucl-th]} \BibitemShut {NoStop}%
	\bibitem [{\citenamefont {Greif}\ \emph {et~al.}(2018)\citenamefont {Greif}, \citenamefont {Fotakis}, \citenamefont {Denicol},\ and\ \citenamefont {Greiner}}]{Greif:2017byw}%
	\BibitemOpen
	\bibfield  {author} {\bibinfo {author} {\bibfnamefont {M.}~\bibnamefont {Greif}}, \bibinfo {author} {\bibfnamefont {J.~A.}\ \bibnamefont {Fotakis}}, \bibinfo {author} {\bibfnamefont {G.~S.}\ \bibnamefont {Denicol}},\ and\ \bibinfo {author} {\bibfnamefont {C.}~\bibnamefont {Greiner}},\ }\bibfield  {title} {\bibinfo {title} {{Diffusion of conserved charges in relativistic heavy ion collisions}},\ }\href {https://doi.org/10.1103/PhysRevLett.120.242301} {\bibfield  {journal} {\bibinfo  {journal} {Phys. Rev. Lett.}\ }\textbf {\bibinfo {volume} {120}},\ \bibinfo {pages} {242301} (\bibinfo {year} {2018})},\ \Eprint {https://arxiv.org/abs/1711.08680} {arXiv:1711.08680 [hep-ph]} \BibitemShut {NoStop}%
	\bibitem [{\citenamefont {Almaalol}\ \emph {et~al.}(2025)\citenamefont {Almaalol}, \citenamefont {Dore},\ and\ \citenamefont {Noronha-Hostler}}]{Almaalol:2022pjc}%
	\BibitemOpen
	\bibfield  {author} {\bibinfo {author} {\bibfnamefont {D.}~\bibnamefont {Almaalol}}, \bibinfo {author} {\bibfnamefont {T.}~\bibnamefont {Dore}},\ and\ \bibinfo {author} {\bibfnamefont {J.}~\bibnamefont {Noronha-Hostler}},\ }\bibfield  {title} {\bibinfo {title} {{Stability of multicomponent relativistic viscous hydrodynamics from Israel-Stewart and reproducing Denicol-Niemi-Molnar-Rischke from maximizing the entropy}},\ }\href {https://doi.org/10.1103/PhysRevD.111.014020} {\bibfield  {journal} {\bibinfo  {journal} {Phys. Rev. D}\ }\textbf {\bibinfo {volume} {111}},\ \bibinfo {pages} {014020} (\bibinfo {year} {2025})},\ \Eprint {https://arxiv.org/abs/2209.11210} {arXiv:2209.11210 [hep-th]} \BibitemShut {NoStop}%
	\bibitem [{\citenamefont {Capellino}\ \emph {et~al.}(2022)\citenamefont {Capellino}, \citenamefont {Beraudo}, \citenamefont {Dubla}, \citenamefont {Floerchinger}, \citenamefont {Masciocchi}, \citenamefont {Pawlowski},\ and\ \citenamefont {Selyuzhenkov}}]{Capellino_2022}%
	\BibitemOpen
	\bibfield  {author} {\bibinfo {author} {\bibfnamefont {F.}~\bibnamefont {Capellino}}, \bibinfo {author} {\bibfnamefont {A.}~\bibnamefont {Beraudo}}, \bibinfo {author} {\bibfnamefont {A.}~\bibnamefont {Dubla}}, \bibinfo {author} {\bibfnamefont {S.}~\bibnamefont {Floerchinger}}, \bibinfo {author} {\bibfnamefont {S.}~\bibnamefont {Masciocchi}}, \bibinfo {author} {\bibfnamefont {J.}~\bibnamefont {Pawlowski}},\ and\ \bibinfo {author} {\bibfnamefont {I.}~\bibnamefont {Selyuzhenkov}},\ }\bibfield  {title} {\bibinfo {title} {Fluid-dynamic approach to heavy-quark diffusion in the quark-gluon plasma},\ }\bibfield  {journal} {\bibinfo  {journal} {Physical Review D}\ }\textbf {\bibinfo {volume} {106}},\ \href {https://doi.org/10.1103/physrevd.106.034021} {10.1103/physrevd.106.034021} (\bibinfo {year} {2022})\BibitemShut {NoStop}%
	\bibitem [{\citenamefont {Capellino}\ \emph {et~al.}(2023)\citenamefont {Capellino}, \citenamefont {Dubla}, \citenamefont {Floerchinger}, \citenamefont {Grossi}, \citenamefont {Kirchner},\ and\ \citenamefont {Masciocchi}}]{Capellino_2023}%
	\BibitemOpen
	\bibfield  {author} {\bibinfo {author} {\bibfnamefont {F.}~\bibnamefont {Capellino}}, \bibinfo {author} {\bibfnamefont {A.}~\bibnamefont {Dubla}}, \bibinfo {author} {\bibfnamefont {S.}~\bibnamefont {Floerchinger}}, \bibinfo {author} {\bibfnamefont {E.}~\bibnamefont {Grossi}}, \bibinfo {author} {\bibfnamefont {A.}~\bibnamefont {Kirchner}},\ and\ \bibinfo {author} {\bibfnamefont {S.}~\bibnamefont {Masciocchi}},\ }\bibfield  {title} {\bibinfo {title} {Fluid dynamics of charm quarks in the quark-gluon plasma},\ }\bibfield  {journal} {\bibinfo  {journal} {Physical Review D}\ }\textbf {\bibinfo {volume} {108}},\ \href {https://doi.org/10.1103/physrevd.108.116011} {10.1103/physrevd.108.116011} (\bibinfo {year} {2023})\BibitemShut {NoStop}%
	\bibitem [{\citenamefont {Hiscock}\ and\ \citenamefont {Olson}(1989)}]{HISCOCK1989125}%
	\BibitemOpen
	\bibfield  {author} {\bibinfo {author} {\bibfnamefont {W.~A.}\ \bibnamefont {Hiscock}}\ and\ \bibinfo {author} {\bibfnamefont {T.~S.}\ \bibnamefont {Olson}},\ }\bibfield  {title} {\bibinfo {title} {Effects of frame choice on nonlinear dynamics in relativistic heat-conducting fluid theories},\ }\href {https://doi.org/https://doi.org/10.1016/0375-9601(89)90772-X} {\bibfield  {journal} {\bibinfo  {journal} {Physics Letters A}\ }\textbf {\bibinfo {volume} {141}},\ \bibinfo {pages} {125} (\bibinfo {year} {1989})}\BibitemShut {NoStop}%
	\bibitem [{\citenamefont {Hiscock}\ and\ \citenamefont {Lindblom}(1988)}]{Hiscock_Lindblom_pathologies_1988}%
	\BibitemOpen
	\bibfield  {author} {\bibinfo {author} {\bibfnamefont {W.~A.}\ \bibnamefont {Hiscock}}\ and\ \bibinfo {author} {\bibfnamefont {L.}~\bibnamefont {Lindblom}},\ }\bibfield  {title} {\bibinfo {title} {Nonlinear pathologies in relativistic heat-conducting fluid theories},\ }\href@noop {} {\bibfield  {journal} {\bibinfo  {journal} {Physics Letters A}\ }\textbf {\bibinfo {volume} {131}},\ \bibinfo {pages} {509} (\bibinfo {year} {1988})}\BibitemShut {NoStop}%
	\bibitem [{\citenamefont {Olson}(1990)}]{Olson:1989ey}%
	\BibitemOpen
	\bibfield  {author} {\bibinfo {author} {\bibfnamefont {T.~S.}\ \bibnamefont {Olson}},\ }\bibfield  {title} {\bibinfo {title} {Stability and causality in the {I}srael-{S}tewart energy frame theory},\ }\href {https://doi.org/10.1016/0003-4916(90)90366-V} {\bibfield  {journal} {\bibinfo  {journal} {Annals Phys.}\ }\textbf {\bibinfo {volume} {199}},\ \bibinfo {pages} {18} (\bibinfo {year} {1990})}\BibitemShut {NoStop}%
	\bibitem [{\citenamefont {Rezzolla}\ and\ \citenamefont {Zanotti}(2013{\natexlab{a}})}]{Rezzolla_Zanotti_book}%
	\BibitemOpen
	\bibfield  {author} {\bibinfo {author} {\bibfnamefont {L.}~\bibnamefont {Rezzolla}}\ and\ \bibinfo {author} {\bibfnamefont {O.}~\bibnamefont {Zanotti}},\ }\href@noop {} {\emph {\bibinfo {title} {Relativistic hydrodynamics}}}\ (\bibinfo  {publisher} {Oxford University Press},\ \bibinfo {address} {New York},\ \bibinfo {year} {2013})\BibitemShut {NoStop}%
	\bibitem [{\citenamefont {Israel}(1976{\natexlab{b}})}]{ISRAEL1976310}%
	\BibitemOpen
	\bibfield  {author} {\bibinfo {author} {\bibfnamefont {W.}~\bibnamefont {Israel}},\ }\bibfield  {title} {\bibinfo {title} {Nonstationary irreversible thermodynamics: A causal relativistic theory},\ }\href {https://doi.org/https://doi.org/10.1016/0003-4916(76)90064-6} {\bibfield  {journal} {\bibinfo  {journal} {Annals of Physics}\ }\textbf {\bibinfo {volume} {100}},\ \bibinfo {pages} {310} (\bibinfo {year} {1976}{\natexlab{b}})}\BibitemShut {NoStop}%
	\bibitem [{\citenamefont {Israel}\ and\ \citenamefont {Stewart}(1979{\natexlab{b}})}]{ISRAEL1979341}%
	\BibitemOpen
	\bibfield  {author} {\bibinfo {author} {\bibfnamefont {W.}~\bibnamefont {Israel}}\ and\ \bibinfo {author} {\bibfnamefont {J.}~\bibnamefont {Stewart}},\ }\bibfield  {title} {\bibinfo {title} {Transient relativistic thermodynamics and kinetic theory},\ }\href {https://doi.org/https://doi.org/10.1016/0003-4916(79)90130-1} {\bibfield  {journal} {\bibinfo  {journal} {Annals of Physics}\ }\textbf {\bibinfo {volume} {118}},\ \bibinfo {pages} {341} (\bibinfo {year} {1979}{\natexlab{b}})}\BibitemShut {NoStop}%
	\bibitem [{\citenamefont {Choquet-Bruhat}(2009)}]{ChoquetBruhatGRBook}%
	\BibitemOpen
	\bibfield  {author} {\bibinfo {author} {\bibfnamefont {Y.}~\bibnamefont {Choquet-Bruhat}},\ }\href@noop {} {\emph {\bibinfo {title} {General Relativity and the Einstein Equations}}}\ (\bibinfo  {publisher} {Oxford University Press},\ \bibinfo {address} {New York},\ \bibinfo {year} {2009})\BibitemShut {NoStop}%
	\bibitem [{\citenamefont {Rees}(1922)}]{Rees_1922}%
	\BibitemOpen
	\bibfield  {author} {\bibinfo {author} {\bibfnamefont {E.~L.}\ \bibnamefont {Rees}},\ }\bibfield  {title} {\bibinfo {title} {Graphical discussion of the roots of a quartic equation},\ }\href {https://doi.org/10.1080/00029890.1922.11986100} {\bibfield  {journal} {\bibinfo  {journal} {The American Mathematical Monthly}\ }\textbf {\bibinfo {volume} {29}},\ \bibinfo {pages} {51} (\bibinfo {year} {1922})},\ \Eprint {https://arxiv.org/abs/https://doi.org/10.1080/00029890.1922.11986100} {https://doi.org/10.1080/00029890.1922.11986100} \BibitemShut {NoStop}%
	\bibitem [{\citenamefont {Rezzolla}\ and\ \citenamefont {Zanotti}(2013{\natexlab{b}})}]{RezzollaZanottiBookRelHydro}%
	\BibitemOpen
	\bibfield  {author} {\bibinfo {author} {\bibfnamefont {L.}~\bibnamefont {Rezzolla}}\ and\ \bibinfo {author} {\bibfnamefont {O.}~\bibnamefont {Zanotti}},\ }\href@noop {} {\emph {\bibinfo {title} {Relativistic Hydrodynamics}}}\ (\bibinfo  {publisher} {Oxford University Press},\ \bibinfo {address} {New York},\ \bibinfo {year} {2013})\BibitemShut {NoStop}%
	\bibitem [{\citenamefont {royale des~sciences (France)}(1835)}]{SturmTheorem}%
	\BibitemOpen
	\bibfield  {author} {\bibinfo {author} {\bibfnamefont {A.}~\bibnamefont {royale des~sciences (France)}},\ }\bibinfo {title} {Mémoires presentés a l'institut des sciences, lettres et arts, par divers savants èt lus dans ses assemblées : Sciences, mathématiques et physiques}\ (\bibinfo {year} {1835})\ p.\ \bibinfo {pages} {276},\ \bibinfo {note} {https://www.biodiversitylibrary.org/bibliography/4363}\BibitemShut {NoStop}%
	\bibitem [{\citenamefont {Kurtz}(1992)}]{SuffCondRealRoots}%
	\BibitemOpen
	\bibfield  {author} {\bibinfo {author} {\bibfnamefont {D.~C.}\ \bibnamefont {Kurtz}},\ }\bibfield  {title} {\bibinfo {title} {A sufficient condition for all the roots of a polynomial to be real},\ }\href {http://www.jstor.org/stable/2325063} {\bibfield  {journal} {\bibinfo  {journal} {The American Mathematical Monthly}\ }\textbf {\bibinfo {volume} {99}},\ \bibinfo {pages} {259} (\bibinfo {year} {1992})}\BibitemShut {NoStop}%
	\bibitem [{\citenamefont {Wald}(2010)}]{WaldBookGR1984}%
	\BibitemOpen
	\bibfield  {author} {\bibinfo {author} {\bibfnamefont {R.~M.}\ \bibnamefont {Wald}},\ }\href@noop {} {\emph {\bibinfo {title} {General relativity}}}\ (\bibinfo  {publisher} {University of Chicago press},\ \bibinfo {year} {2010})\BibitemShut {NoStop}%
	\bibitem [{\citenamefont {Cordeiro}(2025)}]{CordeiroGit2025}%
	\BibitemOpen
	\bibfield  {author} {\bibinfo {author} {\bibfnamefont {I.}~\bibnamefont {Cordeiro}},\ }\href {https://github.com/itc2-github/Nonlinear-causality-of-Israel-Stewart-theory-with-diffusion} {\bibinfo {title} {Nonlinear causality of israel-stewart theory with diffusion}} (\bibinfo {year} {2025}),\ \Eprint {https://arxiv.org/abs/https://github.com/itc2-github/Nonlinear-causality-of-Israel-Stewart-theory-with-diffusion} {https://github.com/itc2-github/Nonlinear-causality-of-Israel-Stewart-theory-with-diffusion} \BibitemShut {NoStop}%
	\bibitem [{\citenamefont {Reula}(2004)}]{ReulaStrongHyperbolic}%
	\BibitemOpen
	\bibfield  {author} {\bibinfo {author} {\bibfnamefont {O.~A.}\ \bibnamefont {Reula}},\ }\bibfield  {title} {\bibinfo {title} {Strongly hyperbolic systems in general relativity},\ }\href {https://doi.org/10.1142/S0219891604000111} {\bibfield  {journal} {\bibinfo  {journal} {J. Hyperbolic Differ. Equ.}\ }\textbf {\bibinfo {volume} {01}},\ \bibinfo {pages} {251} (\bibinfo {year} {2004})}\BibitemShut {NoStop}%
	\bibitem [{\citenamefont {Kontou}\ and\ \citenamefont {Sanders}(2020)}]{Kontou_2020}%
	\BibitemOpen
	\bibfield  {author} {\bibinfo {author} {\bibfnamefont {E.-A.}\ \bibnamefont {Kontou}}\ and\ \bibinfo {author} {\bibfnamefont {K.}~\bibnamefont {Sanders}},\ }\bibfield  {title} {\bibinfo {title} {Energy conditions in general relativity and quantum field theory},\ }\href {https://doi.org/10.1088/1361-6382/ab8fcf} {\bibfield  {journal} {\bibinfo  {journal} {Classical and Quantum Gravity}\ }\textbf {\bibinfo {volume} {37}},\ \bibinfo {pages} {193001} (\bibinfo {year} {2020})}\BibitemShut {NoStop}%
\end{thebibliography}
\end{document}